\documentclass[aps,groupedaddress,preprint,superscriptaddress,showpacs]{revtex4-1}

\usepackage[T1]{fontenc}
\usepackage[latin9]{inputenc}
\usepackage{mathrsfs}
\usepackage{amsmath}
\usepackage{amsfonts}
\usepackage{graphicx}
\usepackage{esint}
\usepackage{fixmath}
\usepackage{wrapfig}
\usepackage{mathrsfs}
\usepackage{setspace}
\usepackage{breakurl}
\usepackage{amsbsy}

\begin{document}

\title{Generation of entangled photons via parametric down-conversion in semiconductor lasers and integrated quantum photonic systems}

\author{ Mikhail Tokman}
\affiliation{Institute of Applied Physics, Russian Academy of Sciences, Nizhny Novgorod, 603950, Russia }
\author{Yongrui Wang}
\affiliation{Department of Physics and Astronomy, Texas A\&M University, College Station, TX, 77843 USA}
\author{Qianfan Chen}
\affiliation{Department of Physics and Astronomy, Texas A\&M University, College Station, TX, 77843 USA}
\author{Leon Shterengas}
\affiliation{State University of New York at Stony Brook, Stony Brook, NY 11794 USA}
\author{Alexey Belyanin}
\affiliation{Department of Physics and Astronomy, Texas A\&M University, College Station, TX, 77843 USA}

\begin{abstract}
We propose and design a high-brightness, ultra-compact electrically pumped GaSb-based laser source of entangled photons generated by mode-matched intracavity parametric down-conversion of lasing modes. To describe the nonlinear mixing in highly dispersive and dissipative waveguides, we develop a nonperturbative quantum theory of parametric down-conversion of waveguide modes which takes into account the effects of modal dispersion, group and phase mismatch, propagation, dissipation, and coupling to noisy reservoirs. We extend our theory to the regime of quantized pump fields with a new approach based on the propagation equation for the state vector which solves the nonperturbative boundary-value problem of the parametric decay of a quantized single-photon pump mode and can be generalized to include the effects of dissipation and noise. Our formalism is applicable to a wide variety of three-wave mixing propagation problems. It provides convenient analytic expressions for interpreting experimental results and predicting the performance of monolithic quantum photonic systems. 
\end{abstract}


\date{\today }

\maketitle
  
\section{Introduction}

The spontaneous parametric down conversion (SPDC) has become a benchmark process for generation of the entangled photon pairs and heralded single photons in the variety of the experiments and for the needs of the rapidly growing field of the quantum information processing; see, e.g., \cite{torres2011,couteau2018} for recent reviews. Typically, the second order nonlinear susceptibility $\chi^{(2)}$ of the birefringent crystals cut at the angle to satisfy the phase matching condition for a specific SPDC process is used; see for instance \cite{kwiat1995}. The utilization of the periodically poled nonlinear crystals making use of the quasi-phase matching led to improved performance. Recent advances in high-quality microcavities, nanoantennas, and metamaterials have led to the prediction \cite{lapine2014,smirnova2016,daroyan2018} and realization (e.g., \cite{furst2010,solntsev2017,marino2019,ma2020}) of compact chip-scale parametric down-conversion sources. An external optical pump is still required in all cases. Strong second-order nonlinearity of III-V semiconductors in combination with their superior 
light emission properties can be used for both generation of the pump light and production of the entangled photon states where the waveguides and other photonic integrated circuit components can be utilized to facilitate phase matching and to perform quantum information processing operations \cite{wang2020}. The monolithic integration of the pump laser and SPDC source on the same platform still remains a holy grail of this technology as its availability will dramatically simplify the experimental arrangements and pave the way to the long-sought scalable approach to quantum sensing, quantum communications and optical quantum computing. It is natural to consider the III-V heterostructures which can produce high power and stable pump lasers and also demonstrate strong second order nonlinearities required for SPDC as a potential platform for such integration. 

In a standard single core laser waveguide heterostructure the phase matching conditions for the efficient SPDC process are virtually impossible to achieve because of normal dispersion: refractive index of all materials away from resonance transitions decreases with wavelength. One approach to defeat the normal dispersion limitation was successfully realized in Bragg-reflection waveguides (BRW) \cite{horn2012} which confine the pump Bragg mode and total internal reflection guided signal and idler modes. The optically pumped BRW devices have been extensively studied \cite{horn2012,kang2016,svozilik2012}. An intracavity non-degenerate optical parametric generation in electrically injected GaAs-based BRW laser was demonstrated in \cite{bijlani2013} and corresponding emitters of the correlated photons generating broadband product states of the near-infrared signal and idler photons were reported \cite{boitier2014}. However, the design complications associated with the presence of Bragg reflectors comprising the device claddings so far led to strong degradation of the near infrared pump laser performance parameters. It is also not trivial to arrange for stable single mode operation of the BRW pump laser for the same reason. These issues were recognized even for near-infrared BRW emitters where the vertical cavity surface emitting laser technology development led to significant advances in design of the Bragg reflector claddings. In GaSb-based lasers capable of intracavity SPDC generation of the mid-infrared (MWIR) correlated photon pairs the situation with BRW is much less developed and it would be extremely challenging to arrange for efficient carrier transport through thick Bragg reflectors comprising the cladding layers of these devices. 

We propose a different approach to achieve phase matching between  $\sim 2$ $\mu$m pump and $\sim 4$ $\mu$m signal and idler waves which is compatible with GaSb-based and any other semiconductor laser technology. It relies on the natural TE-polarized pump mode and utilizes type-II SPDC to produce biphotons and polarization entangled photon pairs as required for quantum technologies. Both degenerate and non-degenerate type II SPDC are possible and, moreover, can be easily selected by tuning the pump wavelength using standard techniques. Our proposed GaSb-based laser heterostructure illustrated in Fig. 1 has a coupled-waveguide design favoring lasing near $\sim 2$ $\mu$m (pump) in TE-polarized asymmetric super mode. The type-II SPDC process will produce entangled photon pairs in TE- and TM-polarized symmetric super modes. The use of the asymmetric super mode allows us to reduce the effective refractive index of the $\sim 2$ $\mu$m pump to achieve efficient phase matching with $\sim 4$ $\mu$m signal and idler symmetric super modes. The device geometry makes maximum use of the large $\chi_{xyz}^{(2)} \sim 100$ pm/V of III-V zinc blende semiconductors for SPDC process, which is significantly higher as compared to conventional nonlinear crystals \cite{shoji1997}. 

Theory of the entangled photon state generation in semiconductor lasers and waveguides has important peculiarities and challenges which have not been addressed  before. First, laser waveguides and other monolithic integrated photonic systems are inherently highly dispersive and dissipative. Including the effects of dispersion, dissipation and noise in a consistent way is crucial for predicting the performance of these devices. For example, we show below that  the quantum noise  can make significant and even dominant contribution within the signal/idler bandwidth even at low ambient temperature. Second, while SPDC has been typically treated as an initial-value problem for the quantized signal and idler fields, the SPDC in a finite-length waveguide presents a boundary-value eigenmode propagation problem affected by phase and group velocity mismatch, dispersion, absorption, and noisy reservoirs. Third, extra challenges arise in the case of a quantized pump field, e.g. a single-photon pump,  as the operator-valued Heisenberg-Langevin equations become nonlinear. While the case of all three quantum fields has been well studied as a mean-field initial-value problem and in the perturbative regime, here  we present a new  approach based on the propagation equation for the state vector which allows us to describe the SPDC process with nonperturbative coupling between quantized single-photon pump, signal, and idler fields as a boundary-value propagation problem.    

The quantum theory of SPDC in monolithic nonlinear waveguides is applicable to any nonlinear propagation problem involving three-wave mixing. It provides convenient analytic expressions for interpreting experimental results.  The proposed device design principles can be applied to a wide variety of III-V semiconductor diode lasers. The specific implementation of the monolithic electrically pumped quantum light source within the III-V-Sb platform offers an extra bonus of covering the MWIR spectral region which holds strong promise for applications in quantum communications, quantum sensing, and imaging. Free space quantum-secured communication links operating in MWIR range offer significant advantages over near-infrared channels due to lower scattering losses. The satellite-based quantum key distribution systems \cite{liao2018} based on near-infrared sources are severely constrained by solar background radiation and, until introduction of the $\lambda \sim 1.5$ $\mu$m quantum light emitters, were restricted to at night operation \cite{liao2017}. The systems operating near 4 $\mu$m will benefit from dramatically reduced solar background and still moderate Earth thermal background \cite{astm,kaushal2017}. The operation at these wavelengths generally improves reliability and throughput of the free space quantum-secured links under adverse weather conditions, scattering and atmospheric turbulence. Recently, the quantum illumination protocols relying on correlations between photons in entangled pairs have been experimentally demonstrated to offer more than an order of magnitude image contrast improvement in the presence of background light, sensor noise, and loss \cite{gregory2020}. 

The modern technology for emitting and detecting MWIR entangled photon pairs is based on multi-wave mixing in free-standing nonlinear optical elements \cite{prab2020,mccracken2018,mancinelli2017}. This puts many exciting applications out of reach. The full integration of the components of the photonic quantum information technology is not yet available even for much more user friendly near-infrared region of spectrum. The development of the quantum information technology in MWIR region of spectrum is even less advanced. For MWIR both generation of the entangled photon pairs and single photon detection require nonlinear converters since even direct single photon counting photodetectors operating in mid-infrared are yet to be developed and realized. The III-V-Sb material system can host electrically injected entangled photon pair emitters together with all other quantum information processing photonic integrated circuit components, thus serving as a common platform for the advancement of MWIR quantum information technology. 

\section{The diode laser design for intracavity type-II SPDC of laser modes}


\begin{figure}[htbp]
\centering\includegraphics[width=13cm]{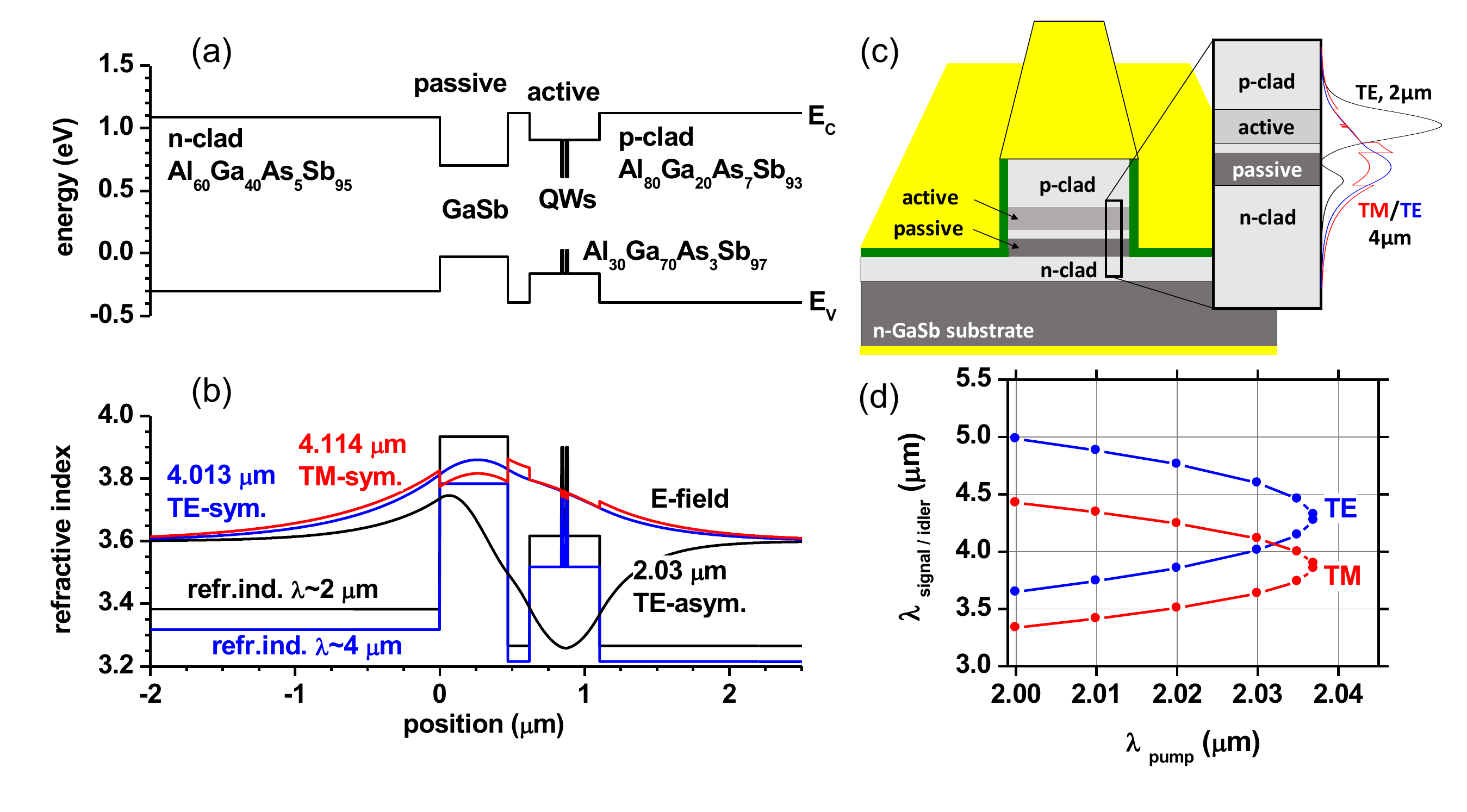}
\caption{ (a) Band diagram of the GaSb-based coupled-waveguide laser heterostructure and (b) refractive index and field profiles (bottom) for the TE polarized 2 $\mu$m pump (blue) and $\sim 3.6$ and $\sim 4.4$ $\mu$m TE (green) and TM (magenta) signal/idler modes. The passive waveguide width is 470 nm and separation barrier between two waveguide cores is 150 nm. The position is along the growth direction. (c) Sketch of a generic laser device with a coupled waveguide for mode-matched Type-II intracavity SPDC of laser photons. The profiles of mode intensities are superimposed on the inset. (d) Calculated wavelengths of the signal and idler modes at exact phase matching as a function of the pump mode wavelength. }
\end{figure}


One example of the laser device design for intracavity type-II SPDC process is shown in Fig.~1. The laser is grown in usual <001> direction and has a ridge cavity aligned along <110> direction; therefore, only the TE $\longrightarrow$ TE+TM SPDC decay is allowed by zinc-blende crystal symmetry. An example of TM and TE signal and idler modes shown in Fig.~1 corresponds to non-degenerate SPDC in which they have different frequencies  and refractive indices. Changing the pump wavelength within the range of $\sim 10$ nm will tune the type-II SPDC from non-degenerate to degenerate case. Fig.~1(d) plots the wavelengths of the phase-matched signal and idler TM and TE polarized photons in symmetric super modes versus wavelength of the TE polarized pump in asymmetric supermode calculated for the coupled waveguide in Fig.~1(b).

The tuning curves in Fig.~1  correspond to perfect phase-matching conditions. The phase matching lines are significantly broadened and the SPDC proceeds within a broad bandwidth; see a detailed discussion below. Due to the broadening the output of the close-to-degenerate SPDC process can be split into many channels with the help of external or integrated spectral filters. 
The inherent possibility of the proposed device heterostructure to achieve and maintain the phase matching conditions while tuning the pump wavelength in a relatively narrow range is one of the key advantages of the proposed design making it a practical solution for the development of the robust and efficient entangled photon pair emitters. 


\section{Initial-value problem for spontaneous parametric down-conversion}

Although our ultimate goal is to solve the boundary-value propagation
problem for coupled pump, signal, and idler modes in a finite-length laser
waveguide, to clarify some universal properties of SPDC here we outline the solution of the
initial-value problem, which is much better studied. This will allow us to see which degrees of freedom can be entangled in the intracavity SPDC process, and which ones cannot. The initial-value problem describes SPDC of cavity modes \cite{daroyan2018,tokman2019}, especially in high-Q cavities, although in our case the laser waveguide is long and lossy enough for the parametric decay to develop in a single-pass propagation regime. 

Consider a dispersive and anisotropic but
uniform nonlinear medium of volume $V$. It is described by the linear
permittivity tensor $\overleftrightarrow{\varepsilon }\left( \omega \right) $
and the second-order (rank 3) nonlinear susceptibility tensor $%
\overleftrightarrow{\overleftrightarrow{\chi }}^{\left( 2\right) }\left(
\omega _{1}+\omega _{2}=\omega _{3}\right) $. We are using this double-arrow notation for dielectric response tensors to save the hat notation for quantum-mechanical operators.

Consider the fields at fixed frequencies satisfying the energy conservation
in the parametric decay,
\begin{equation}
\omega _{p}=\omega_{V}+\omega_{H},  \label{energy conservation}
\end{equation}%
where the electric field at frequency $\omega _{p}$ is a classical pump field, $
\mathbf{E}_{p}e^{-i\omega _{p}t+i\mathbf{k}_{p}\cdot
\mathbf{r}}+$ c.c.  
The Schr\"{o}dinger operators of the quantum field at frequencies $\omega _{V,H}$
are defined as \cite{fain1969,tokman2015}%
\begin{equation}
\mathbf{\hat{E}=}\sum_{\mathbf{k}}\left( \hat{c}_{V\mathbf{k}}\mathbf{E}_{V%
\mathbf{k}}e^{i\mathbf{k\cdot r}}+h.c.\right) +\sum_{\mathbf{q}}\left( \hat{c%
}_{H\mathbf{q}}\mathbf{E}_{H\mathbf{q}}e^{i\mathbf{q\cdot r}}+h.c.\right) .
\label{Schrodinger operator}
\end{equation}%
Here $\hat{c}_{V\mathbf{k}}$ and $\hat{c}_{H\mathbf{q}}$ are standard
bosonic annihilation operators; wave vectors $\mathbf{k}\left( \omega
_{V}\right) $ and $\mathbf{q}\left( \omega _{H}\right) $ are determined from
the dispersion relations for the eigenwaves with
periodic boundary conditions; $\mathbf{E}_{V\mathbf{k}}$ and $\mathbf{E}_{H%
\mathbf{q}}$ are normalization amplitudes for the fields. The direction of $%
\mathbf{E}_{V\mathbf{k}}$ and $\mathbf{E}_{H\mathbf{q}}$ correspond to the
polarization of the eigenmodes in the medium with dielectric permittivity
tensor $\overleftrightarrow{\varepsilon }\left( \omega \right) $. We denote the polarizations by indices $V$ (vertical) and $H$
(horizontal), although the polarization state of the eigenmodes in an
anisotropic medium can be more complex \cite{ginzburg1966}.

The dispersion equations and polarizations of the normalized field
amplitudes in Eq.~(\ref{Schrodinger operator}) are found by solving a
classical electrodynamics problem, whereas the \textit{magnitudes} of these
vector amplitudes have to be determined by field quantization in the volume $%
V$ \cite{fain1969,tokman2015}:%
\begin{equation}
\mathbf{E}_{V\mathbf{k}}^{\ast }\left\{ \frac{\partial \left[ \omega ^{2}%
\overleftrightarrow{\varepsilon }\left( \omega \right) \right] }{\omega
\partial \omega }\right\} _{\omega =\omega _{V}}\mathbf{E}_{V\mathbf{k}}=%
\frac{4\pi \hbar \omega _{V}}{V},\ \ \ \mathbf{E}_{H\mathbf{q}}^{\ast
}\left\{ \frac{\partial \left[ \omega ^{2}\overleftrightarrow{\varepsilon }%
\left( \omega \right) \right] }{\omega \partial \omega }\right\} _{\omega
=\omega _{H}}\mathbf{E}_{H\mathbf{q}}=\frac{4\pi \hbar \omega _{H}}{V}.
\label{field quantization}
\end{equation}

The second-order nonlinearity gives rise to three-wave mixing. Exactly at 
resonance described by Eq.~(\ref{energy conservation}) the Hamiltonian of
the system in the interaction picture and rotating-wave approximation (RWA)
takes the form (see, e.g., \cite{morton1994})%
\begin{equation}
\hat{H}_{int}=-\sum_{\mathbf{k,q}}\left( M_{\mathbf{kq}}\hat{c}_{V\mathbf{k}%
}^{\dagger }\hat{c}_{H\mathbf{q}}^{\dagger }+h.c.\right) ,
\label{Hamiltonian}
\end{equation}%
where%
\begin{equation}
M_{\mathbf{kq}}=\mathbf{E}_{V\mathbf{k}}^{\ast }\overleftrightarrow{%
\overleftrightarrow{\chi }}^{\left( 2\right) }\left( \omega _{p}-\omega
_{H}=\omega _{V}\right) \mathbf{E}_{p}\ \mathbf{E}_{H\mathbf{q}%
}^{\ast }\int_{V}e^{i\left( \mathbf{k}_{p}-\mathbf{k}\left( \omega
_{V}\right) -\mathbf{q}\left( \omega _{H}\right) \right) \mathbf{\cdot r}%
}d^{3}r,  \label{mkq}
\end{equation}%
and%
\begin{equation}
\mathbf{E}_{V\mathbf{k}}^{\ast }\overleftrightarrow{\overleftrightarrow{\chi
}}^{\left( 2\right) }\left( \omega _{p}-\omega _{H}=\omega _{V}\right) %
\mathbf{E}_{p}\ \mathbf{E}_{H\mathbf{q}}^{\ast }=\mathbf{E}_{H\mathbf{%
q}}^{\ast }\overleftrightarrow{\overleftrightarrow{\chi }}^{\left(
2\right) }\left( \omega _{p}-\omega _{V}=\omega _{H}\right) \mathbf{E}_{p}\ \mathbf{E}_{V\mathbf{k}}^{\ast }.  \label{relationship}
\end{equation}%
(for the last relationship see, e.g., Chapter 2.9 in \cite{keldysh1994}).

Now assume that there is only one wave vector $\mathbf{k}$ for each wave
vector $\mathbf{q}$ (and vice versa) in the sum $\sum_{\mathbf{k,q}}\left(
\cdots \right) $ in Eq.~(\ref{Hamiltonian}) for which the phase matching
condition $\mathbf{k}_{p}\left( \omega _{p}\right) =\mathbf{k}\left( \omega
_{V}\right) +\mathbf{q}\left( \omega _{H}\right) $, the energy conservation
Eq.~(\ref{energy conservation}), and polarization selection rules imposed by
the nonlinear susceptibility tensor are satisfied simultaneously. In this
case the Hamiltonian of the system can be written as%
\begin{equation}
\hat{H}_{int}=\sum_{\mathbf{k,q}}\hat{H}_{\mathbf{kq}}\left( \hat{c}_{V%
\mathbf{k}}^{\dagger },\hat{c}_{V\mathbf{k}},\hat{c}_{H\mathbf{q}}^{\dagger
},\hat{c}_{H\mathbf{q}}\right) ,  \label{Hamiltonian of the system}
\end{equation}%
In which any two terms $\hat{H}_{\mathbf{kq}}$ and $\hat{H}_{\mathbf{k}%
^{\prime }\mathbf{q}^{\prime }}$ of the sum have \textit{no common operator}%
. The operator $\hat{H}_{\mathbf{kq}}$ does not act on the state $\left\vert
\Psi _{V\mathbf{k}^{\prime };H\mathbf{q}^{\prime }}\right\rangle $ , if the
pairs of vectors $\mathbf{k},\mathbf{q}$ and $\mathbf{k}^{\prime },\mathbf{q}%
^{\prime }$ are different. Here the notation $\left\vert \Psi _{V\mathbf{k};H%
\mathbf{q}}\right\rangle $ corresponds to the state vector for two degrees
of freedom of the field: the $V$-mode with wave vector $\mathbf{k}$ and the $%
H$-mode with wave vector $\mathbf{q}$, or signal and idler in the SPDC
process.

It is easy to show (see, e.g., \cite{tokman20152} and the Appendix) by solving the Schr\"{o}dinger equation $i\hbar \frac{\partial }{\partial t}\left\vert \Psi
\right\rangle =\hat{H}_{int}\left\vert \Psi \right\rangle $ with the
Hamiltonian (\ref{Hamiltonian of the system}) that 
if the state vector was in a factorized form at the initial moment of time, i.e., 
$
\left\vert \Psi \left( t=0\right) \right\rangle =\prod_{\mathbf{k,q}%
}\left\vert \Psi _{V\mathbf{k};H\mathbf{q}}\left( t=0\right) \right\rangle $, 
it preserves the factorized
form: 
\begin{equation}
\left\vert \Psi(t) \right\rangle =\prod_{\mathbf{k,q}} e^{-\frac{i}{\hbar }\hat{H}_{\mathbf{kq}}t}\left\vert \Psi _{V\mathbf{k};H%
\mathbf{q}}\left( t=0\right) \right\rangle .
\label{Solution}
\end{equation}
Therefore, the state vector for each pair of the signal and idler degrees of
freedom $V\mathbf{k}$ and $H\mathbf{q}$ corresponds to their entangled state
due to the nature of the SPDC, but any such pair is not entangled with any
other pair.

For the vacuum
initial state $\prod_{\mathbf{k,q}}\left\vert 0_{V\mathbf{k}}\right\rangle
\left\vert 0_{H\mathbf{q}}\right\rangle $ one obtains
\begin{equation}
\left\vert \Psi \right\rangle =\prod_{\mathbf{k,q}}\sum_{n=0}^{\infty }\frac{%
1}{n!}\left( \frac{it_{int}}{\hbar }\right) ^{n}\left( M_{\mathbf{kq}}\hat{c}%
_{V\mathbf{k}}^{\dagger }\hat{c}_{H\mathbf{q}}^{\dagger }+M_{\mathbf{kq}%
}^{\ast }\hat{c}_{V\mathbf{k}}\hat{c}_{H\mathbf{q}}\right) ^{n}\left\vert
0_{V\mathbf{k}}\right\rangle \left\vert 0_{H\mathbf{q}}\right\rangle ,
\label{psi}
\end{equation}%
where $\left\vert 0_{V\mathbf{k}}\right\rangle ,\left\vert 0_{H\mathbf{q}%
}\right\rangle $ are vacuum states for the corresponding degrees of freedom
and $t_{int}$ is the characteristic time of the SPDC development determined
by the interaction length. Within a pure initial-value problem, the value of
$t_{int}$ cannot be calculated and has to be estimated from some ad hoc
considerations. For example if the propagation is along $z$-axis, the
characteristic time can be estimated as $t_{int}\approx \frac{L_{z}}{\sqrt{%
\upsilon_{V_{z}}\upsilon_{H_{z}}}}$, where $\upsilon_{V_{z}}$, $\upsilon_{H_{z}}$ are the group
velocities of the eigenmodes along $z$ and $L_{z}$ is the propagation
length; see the Appendix. 

One can get an important insight into the nature of the SPDC state by
comparing Eq.~(\ref{psi}), which is an exact solution to the Schr\"{o}dinger
equation with the popular expression obtained by the perturbation expansion
in the linear approximation with respect to the interaction Hamiltonian \cite{morton1994,couteau2018},
\begin{equation}
\left\vert \Psi \right\rangle =\left\vert 0\right\rangle -\frac{i}{\hbar} \int_{0}^{t}\hat{H}%
_{int}dt\left\vert 0\right\rangle  \label{exact solution}
\end{equation}%
where $\left\vert 0\right\rangle \equiv \prod_{\mathbf{k,q}}\left\vert 0_{V%
\mathbf{k}}\right\rangle \left\vert 0_{H\mathbf{q}}\right\rangle $. As a
simple illustration, consider the degenerate case \cite%
{couteau2018,kwiat1995} when $\omega _{V}=\omega _{H}=\omega _{p}/2$ and the
sum in Eq.~(\ref{Hamiltonian}) contains only two pairs of the wave vectors: $%
\mathbf{k=k}_{1}$, $\mathbf{q=k}_{2}$ and $\mathbf{k=k}_{2}$, $\mathbf{q=k}%
_{1}$. Denoting the states corresponding to wave vectors $\mathbf{k}_{1}$
and $\mathbf{k}_{2}$ as 1 and 2, the perturbative solution in Eq.~(\ref%
{exact solution}) becomes 
\begin{eqnarray}
\left\vert \Psi \right\rangle &=& \left\vert 0_{V1}\right\rangle \left\vert
0_{H2}\right\rangle \left\vert 0_{V2}\right\rangle \left\vert
0_{H1}\right\rangle  \nonumber \\
&+& \frac{i}{\hbar }t_{int}\left( M_{12}\left\vert
1_{V1}\right\rangle \left\vert 1_{H2}\right\rangle \left\vert
0_{V2}\right\rangle \left\vert 0_{H1}\right\rangle +M_{21}\left\vert
1_{V2}\right\rangle \left\vert 1_{H1}\right\rangle \left\vert
0_{V1}\right\rangle \left\vert 0_{H2}\right\rangle \right) ,
\label{perturbative solution}
\end{eqnarray}%
where $\left\vert 1_{\cdots }\right\rangle $ are single-photon states.
Products $\left\vert 1_{V1}\right\rangle \left\vert 1_{H2}\right\rangle $
and $\left\vert 1_{V2}\right\rangle \left\vert 1_{H1}\right\rangle $
correspond obviously to different pairs of degrees of freedom.

When interpreting the experimental results in which only the photon fluxes
are detected, the vacuum term $\left\vert 0_{V1}\right\rangle \left\vert
0_{H2}\right\rangle \left\vert 0_{V2}\right\rangle \left\vert
0_{H1}\right\rangle $ in the state vector can be omitted, since it is not
observable. Note that the vacuum term cannot be dropped when interpreting the results of the heterodyning experiments,  such as those used to detect a squeezed vacuum state, see,
e.g., \cite{tokman2013}.

After dropping the vacuum state and introducing 
``intuitive'' (although inaccurate) notations $\left\vert
1_{V1}\right\rangle \left\vert 0_{H1}\right\rangle \Rightarrow \left\vert
V_{1}\right\rangle $, $\left\vert 1_{H1}\right\rangle \left\vert
0_{V1}\right\rangle \Rightarrow \left\vert H_{1}\right\rangle $ etc., Eq.~(%
\ref{perturbative solution}) is often written in the following form (see,
e.g., \cite{couteau2018,kwiat1995}):%
\begin{equation}
\left\vert \Psi \right\rangle \propto \left\vert V_{1}\right\rangle
\left\vert H_{2}\right\rangle +e^{i\varphi }\left\vert V_{2}\right\rangle
\left\vert H_{1}\right\rangle ,  \label{ps}
\end{equation}%
where $M_{12}e^{i\varphi }=M_{21}$. The state vector in the form of Eq.~(\ref%
{ps}) is often used as an illustration of the entangled state generation via
SPDC. However, it is important to keep in mind that Eq.~(\ref{ps}) is only
the first order in the perturbation expansion, whereas an \textit{exact}
solution in Eq.~(\ref{Solution}) gives the state vector as 
\begin{eqnarray*}
\left\vert \Psi_{V1;H2;V2;H1}\right\rangle &=& \left\vert \Psi
_{V1;H2}\right\rangle \left\vert \Psi _{V2;H1}\right\rangle \\
&=& \left(
\sum_{n=0}^{\infty }C_{n\left( V1\right) \left( H2\right) }\left\vert
n_{_{V1}}\right\rangle \left\vert n_{H2}\right\rangle \right) \times \left(
\sum_{m=0}^{\infty }C_{m\left( V2\right) \left( H1\right) }\left\vert
m_{V1}\right\rangle \left\vert m_{H2}\right\rangle \right),
\end{eqnarray*}%
where $\left\vert n_{\cdots }\right\rangle $ and $\left\vert m_{\cdots
}\right\rangle $ are Fock states. One can see from the exact solution that the
entanglement takes place only within each pair of the degrees of freedom $%
V1\Leftrightarrow H2$ and $V2\Leftrightarrow H1$ whereas the states of
\textit{different} pairs $\left\vert \Psi _{V1;H2}\right\rangle $ and $%
\left\vert \Psi _{V2;H1}\right\rangle $ cannot be entangled. In interpreting specific experiments, one can deal with states like the one in Eq.~(\ref{ps}) in the above approximate sense and after applying spatial, spectral, and time-bin selection designed to avoid collecting (or to ``erase'') certain information. Of course one can also create polarization-entangled states of the type in Eq.~(\ref{ps}) by projecting the original biphoton product state onto a different polarization basis using external optical elements; see the theory in \cite{morton1994}. 

The above analysis is true only for a classical pump. The decay of the
photons of a \textit{quantum} field at frequency $\omega _{p}$ leads to a
complete entanglement of all degrees of freedom; see Sec. 5 and the Appendix.


\section{Finite waveguide: the boundary-value problem for Heisenberg
operators}

Now that we reminded the reader of the nature of biphoton states generated in SPDC, we
can move closer to the waveguide propagation problem of the parametric decay
of a given laser mode. Consider the field propagating along $z$ axis, with
the waveguide cross-section of the total area $S$ in the $x,y$ plane. To
calculate the generation rate of two-photon states, we again assume that the
laser mode is described by a classical coherent field (the pump),
$$ \mathbf{E}_{p}\left( \mathbf{r}_{\perp
}\right) e^{ik_{p}z-i\omega _{p}t}+\mathbf{E}_{p}^{\ast }\left(
\mathbf{r}_{\perp }\right) e^{-ik_{p}z+i\omega _{p}t},
$$
where $\mathbf{r}_{\perp }=\left( x,y\right) .$

The quantized waveguide modes within each pair of decay photons have to be
of different polarizations, \textit{TE} and \textit{TM} type, and satisfy
the energy conservation similar to Eq.~(\ref{energy conservation}),
\begin{equation}
\omega _{p}=\omega _{TE}+\omega _{TM}.  \label{ec}
\end{equation}

In the boundary-value waveguide propagation problem we need to describe the
quantized field of decay photons with time- and coordinate-dependent field
operators. We will introduce the mode index $N=TE,TM$ for brevity, which
labels both the field polarization and the transverse profile of the field $%
\mathbf{E}_{N}\left( \mathbf{r}_{\perp }\right) $. Its dispersion equation
is $\omega =\omega _{N}\left( k\right) $. It is convenient to set apart fast
space-time oscillations at the optical frequency and wave number, and
introduce operators associated with slowly varying field amplitudes:%
\begin{equation}
\mathbf{\hat{E}}_{N}\mathbf{=}\hat{c}_{N}\left( z,t\right) \mathbf{E}%
_{N}\left( \mathbf{r}_{\perp }\right) e^{ik_{N}\left( \omega _{N}\right)
z-i\omega _{N}t}+\hat{c}_{N}^{\dagger }\left( z,t\right) \mathbf{E}%
_{N}^{\ast }\left( \mathbf{r}_{\perp }\right) e^{-ik_{N}\left( \omega
_{N}\right) z+i\omega _{N}t}.  \label{slowly varying field}
\end{equation}%
The normalization of the field is
\begin{equation}
\int_{S}\mathbf{E}_{N}^{\ast }\left( \mathbf{r}_{\perp }\right) \left\{
\frac{\partial \left[ \omega ^{2}\overleftrightarrow{\varepsilon }\left(
\omega ,\mathbf{r}_{\perp }\right) \right] }{\omega \partial \omega }%
\right\} _{\omega =\omega _{N}}\mathbf{E}_{N}\left( \mathbf{r}_{\perp
}\right) d^{2}r=4\pi \hbar \omega _{N},  \label{normalization of the field}
\end{equation}%
where $\overleftrightarrow{\varepsilon }\left( \omega ,\mathbf{r}_{\perp
}\right) $ is the linear dielectric permittivity tensor. With this
definition the dyadic $\hat{c}_{N}^{\dagger }\left( z,t\right) \hat{c}%
_{N}\left( z,t\right) $ is the operator of a photon number per unit length
along $z$, which can slowly change with time and $z$. 

The operators $\hat{c}_{N}^{\dagger }\left( z,t\right) ,\hat{c}_{N}\left(
z,t\right) $ introduced this way obey the commutation relations (see \cite%
{tokman20152,tokman2013,vdovin2013,sisakyan2007,tokman2016,erukhimova2017})
\begin{equation}
\left[ \hat{c}_{N\nu }\left( z\right) ,\hat{c}_{N^{\prime }\nu ^{\prime
}}^{\dagger }\left( z\right) \right] =\delta _{NN^{\prime }}\frac{\delta
\left( \nu -\nu ^{\prime }\right) }{2\pi \upsilon_{N}},
\label{commutation relations}
\end{equation}%
where%
\begin{equation}
\hat{c}_{N}\left( z,t\right) =\int_{\Delta \omega }\hat{c}_{N\nu }\left(
z\right) e^{-i\nu t}d\nu ,\ \ \hat{c}_{N}^{\dagger }\left( z,t\right)
=\int_{\Delta \omega }\hat{c}_{N\nu }^{\dagger }\left( z\right) e^{i\nu
t}d\nu ,  \label{operators}
\end{equation}%
$\Delta \omega$ is the frequency bandwidth occupied by the quantized field and  $\upsilon_{N}=\frac{\partial \omega _{N}}{\partial k}$is the group velocity. The
factor $\frac{1}{2\pi \upsilon_{N}}$ comes from the density of states argument and
corresponds to the ratio $\frac{\Delta n_{N}}{L}$, where $\Delta n_{N}$ is
the number of states in the interval $d\omega $ when a given mode with index
$N$ is quantized within a segment $L$ with periodic boundary conditions. Eqs.~(\ref{commutation relations}) and (\ref{operators}) reflect the fact that the field envelopes occupy a narrow but finite
bandwidth.  

In this section we use the Heisenberg-Langevin formalism to calculate the evolution of
the field operators.  We will follow our previous work \cite%
{tokman20152,tokman2013,vdovin2013,tokman2016,erukhimova2017,tokman2018,tokman2019}. The step-by-step derivation for the general non-degenerate SPDC process is in the Appendix. Here we consider only the degenerate SPDC case 
 when $\omega _{TE}=\omega _{TM}=\omega
_{p}/2$. We assume that exact phase matching is reached for central
frequencies, $k_{TM}\left( \frac{\omega _{p}}{2}\right) +k_{TE}\left( \frac{%
\omega _{p}}{2}\right) -k_{p}=0$. The phase mismatch still accumulates with finite detuning $\nu$ from the central frequencies, determining the SPDC bandwidth as we see below. Generalizing to an arbitrary phase mismatch and non-degenerate SPDC is straightforward but more cumbersome  and the general result is in the Appendix. 


\subsection{Heisenberg-Langevin equations for field operators}

The coupled
equations for the slowly varying field operators are 
\begin{equation}
\left( \frac{\partial }{\partial t}+ \Gamma_{TE} +\upsilon_{TE}\frac{\partial }{\partial z} \right)
\hat{c}_{TE} - \frac{i}{\hbar }A\hat{c}_{TM}^{\dagger }  = \hat{L}_{TE},
\label{ceq for cte}
\end{equation}%
\begin{equation}
\left( \frac{\partial }{\partial t} + \Gamma_{TM} +\upsilon_{TM}\frac{\partial }{%
\partial z} \right) \hat{c}_{TM}^{\dagger } + \frac{i}{\hbar }A^{\ast }\hat{c}_{TE}  = \hat{L}^{\dagger}_{TM}.
\label{ceq for ctm}
\end{equation}
Here 
\begin{equation}
A=\int_{S}\mathbf{E}_{TE}^{\ast }\left( \mathbf{r}_{\perp }\right) \left[
\overleftrightarrow{\overleftrightarrow{\chi }}^{\left( 2\right) }\left(
\mathbf{r}_{\perp }\right) \mathbf{E}_{p}\left( \mathbf{r}_{\perp
}\right) \mathbf{E}_{TM}^{\ast }\left( \mathbf{r}_{\perp }\right) \right]
d^{2}r,  \label{A}
\end{equation}%
where $\overleftrightarrow{\overleftrightarrow{\chi }}^{\left( 2\right) }$
is the second-order nonlinear susceptibility, and $\mathbf{E}_{TE}^{\ast }%
\overleftrightarrow{\overleftrightarrow{\chi }}_{p}^{\left( 2\right) }%
\mathbf{E}_{p}\mathbf{E}_{TM}^{\ast }=\mathbf{E}_{TM}^{\ast }%
\overleftrightarrow{\overleftrightarrow{\chi }}_{p}^{\left( 2\right) }%
\mathbf{E}_{p}\mathbf{E}_{TE}^{\ast }$ \cite{keldysh1994}.
The factors $\Gamma _{N}$ determine modal
losses for the field and are related to the Langevin noise operators $\hat{L}_{N}$ through
fluctuation-dissipation relations (see \cite{tokman20152,tokman2013,vdovin2013,tokman2016,erukhimova2017,tokman2018}). Following \cite{vdovin2013,tokman2016,erukhimova2017}, we
will use the following relationships for the Langevin noise operators, 
\begin{equation}
\left[ \hat{L}_{N\nu }\left( z\right) ,\hat{L}_{N^{\prime }\nu ^{\prime
}}^{\dagger }\left( z^{\prime }\right) \right] =\frac{\Gamma _{N}}{\pi }%
\delta _{NN^{\prime }}\delta \left( \nu -\nu ^{\prime }\right) \delta \left(
z-z^{\prime }\right) ,  \label{commutator for L}
\end{equation}%
\begin{equation}
\left\langle \hat{L}^{\dagger}_{N\nu }\left( z\right) \hat{L}_{N^{\prime }\nu^{\prime
}}\left( z^{\prime }\right) \right\rangle =\frac{\Gamma
_{N}n_{T}\left( \omega _{N}\right) }{\pi }\delta _{NN^{\prime }}\delta
\left( \nu -\nu ^{\prime }\right) \delta \left( z-z^{\prime }\right) ,
\label{cf for L}
\end{equation}%
where $\left\langle \cdots \right\rangle $ means averaging over both an
initial quantum state in the Heisenberg picture and the statistics of the
dissipative reservoir, $n_{T}\left( \omega \right) =\left( e^{\hbar \omega
/T}-1\right) ^{-1}$,
\begin{equation*}
\hat{L}_{N}=\int_{\Delta \omega }\hat{L}_{N\nu }e^{-i\nu t}d\nu ,\ \ \hat{L}%
_{N}^{\dagger }=\int_{\Delta \omega }\hat{L}_{N\nu }^{\dagger }e^{i\nu
t}d\nu .
\end{equation*}%
Equation~(\ref{commutator for L}) ensures the conservation of the commutation relation Eq.~(\ref{commutation relations}) despite the presence of dissipation.

Equations~(\ref{ceq for cte}) and (\ref{ceq for ctm}) have the boundary conditions%
\begin{equation}
\hat{c}_{N}\left( t,z=0\right) =\hat{c}_{N}^{\left( 0\right) }\left(
t\right) .  \label{boundary conditions}
\end{equation}%
The slow time dependence in $\hat{c}_{N}^{\left( 0\right) }$ is due to a
finite (although narrow) bandwidth $\Delta \omega $,%
\begin{equation}
\hat{c}_{N}^{\left( 0\right) }\left( t\right) =\int_{\Delta \omega }\hat{c}%
_{N\nu }^{\left( 0\right) }e^{-i\nu t}d\nu ,\ \ \hat{c}_{N}^{\left( 0\right)
\dagger }\left( t\right) =\int_{\Delta \omega }\hat{c}_{N\nu }^{\left(
0\right) \dagger }e^{i\nu t}d\nu ,  \label{slow time dependence}
\end{equation}%
where $\hat{c}_{N\nu }^{\left( 0\right) }$ is the Schr\"{o}dinger (constant)
operator. If the field at the boundary is an incoherent noise field with a
certain spectral photon distribution $n\left( \omega \right) $ , the
following useful relationships are satisfied:
\begin{equation}
\left\langle \hat{c}_{N\nu }^{\left( 0\right) \dagger }\hat{c}_{N^{\prime
}\nu ^{\prime }}^{\left( 0\right) }\right\rangle =n\left( \omega _{N}\right)
\delta _{NN^{\prime }}\frac{\delta \left( \nu -\nu ^{\prime }\right) }{2\pi
\upsilon_{N}},\ \ \ \left\langle \hat{c}_{N\nu }^{\left( 0\right) }\hat{c}%
_{N^{\prime }\nu ^{\prime }}^{\left( 0\right) \dagger }\right\rangle =\left[
n\left( \omega _{N}\right) +1\right] \delta _{NN^{\prime }}\frac{\delta
\left( \nu -\nu ^{\prime }\right) }{2\pi \upsilon_{N}}  \label{relationships}
\end{equation}%
The  photon flux in
the narrow frequency band $\Delta \omega $ is $Q_{N}=\upsilon_{N}\left\langle \hat{c%
}_{N}^{\left( 0\right) \dagger }\hat{c}_{N}^{\left( 0\right) }\right\rangle
=n\left( \omega _{N}\right) \frac{\Delta \omega }{2\pi }$. In particular, for vacuum boundary conditions in Eq.~(\ref{relationships})
we have $n\left( \omega _{N}\right) =0$. For a thermal
noise we have $n_{T}\left( \omega_N \right) =\left( e^{\hbar \omega_N 
/T}-1\right) ^{-1}$, where $T$ is temperature in energy units.

In the boundary-value problem, it is convenient to transfer from the
operators $\hat{c}_{N }$ which
determine the \textit{ density} of the photon number per unit length
along the waveguide, $\left\langle \hat{c}_{N }^{\dagger }\hat{c}%
_{N }\right\rangle $, to the operators $\hat{a}_{N }=\sqrt{\upsilon_{N}}\hat{c%
}_{N }$  which determine the \textit{flux} of photons in
the waveguide, $\left\langle \hat{a}_{N }^{\dagger }\hat{a}%
_{N }\right\rangle $. 

Next, we transfer to the flux operators in Eqs.~(\ref{ceq for cte}) and (\ref{ceq
for ctm}) and use the Fourier expansion 
\begin{equation}
\hat{a}_{N}\left( z,t\right) =\int_{\Delta \omega }\hat{a}%
_{N\nu }e^{-i\nu t}d\nu ,\ \ \hat{a}_{N}^{
\dagger }\left( z,t\right) =\int_{\Delta \omega }\hat{a}_{N\nu }^{ \dagger }e^{i\nu t}d\nu.   \label{anu}
\end{equation}
The flux operators $\hat{a}_{N\nu }$ satisfy the
commutation relations that follow from Eq.~(\ref{commutation relations}),
namely%
\begin{equation}
\left[ \hat{a}_{N\nu}\left( z\right) ,\hat{a}_{N^{\prime }\nu
^{\prime }}^{\dagger }\left( z\right) \right] =\delta _{NN^{\prime }}%
\frac{\delta \left( \nu-\nu ^{\prime }\right) }{2\pi }.
\label{cr}
\end{equation}
This gives 
\begin{equation}
\left( -i\frac{\nu + i \Gamma_{TE} }{\upsilon_{TE}}  +\frac{\partial }{\partial z} \right) \hat{a}_{TE\nu } - ig\hat{a}_{TM\left( -\nu \right) }^{\dagger } =\frac{1}{\sqrt{\upsilon_{TE}}} \hat{L}_{TE\nu}(z)  ,
\label{rhomb1}
\end{equation}
\begin{equation}
\left( -i\frac{\nu + i \Gamma_{TM} }{\upsilon_{TM}}  +\frac{%
\partial }{\partial z} \right) \hat{a}_{TM\left( -\nu \right) }^{\dagger } + ig^{\ast }%
\hat{a}_{TE\nu } = \frac{1}{\sqrt{\upsilon_{TM}}} \hat{L}^{\dagger}_{TM(-\nu)}(z),  \label{rhomb2}
\end{equation}
where the coupling coefficient 
\begin{equation}
g=\frac{A}{\hbar \sqrt{\upsilon_{TE}\upsilon_{TM}}}.  \label{g}
\end{equation}


\subsection{Observable biphoton fluxes }

The solution for operators $\hat{a}_{N\nu }$ is given in the Appendix (see similar derivations in \cite{tokman2013,vdovin2013,erukhimova2017}). 
Here we give the final expressions for the observable spectral fluxes of photons at the cross section $z = L$ of the waveguide. 
In the absence of coherent incident fields at signal and idler frequencies we have $\left\langle  \hat{a}_{N\nu}^{\dagger }\left( z\right) \hat{a}_{N\nu'}\left( z\right) \right\rangle \propto \delta(\nu - \nu')$. 
Using the solution for the flux operators from the Appendix for vacuum boundary conditions and Langevin noise given by Eqs.~(\ref{commutator for L}) and (\ref{cf for L}), we arrive at
\begin{equation}
Q_{N\nu}(L) = \int_{\Delta \omega} d\nu' \left\langle  \hat{a}_{N\nu}^{\dagger }\left( L\right) \hat{a}_{N\nu'}\left( L\right) \right\rangle  = Q^{(s)}_{N\nu}(L) + Q_{N\nu}^{\rm noise}(L), 
\label{qnu4}
\end{equation}
where we separated the ``signal'' component of the flux $Q^{(s)}_{N\nu}$ and the noise component $Q_{N\nu}^{\rm noise}$ which does not depend on the boundary conditions for the fields: 
\begin{equation}
Q^{(s)}_{TE\nu}(L) = Q^{(s)}_{TM(-\nu)}(L) = e^{-\left( 
\frac{\Gamma_{TE}}{\upsilon_{TE}} + \frac{\Gamma_{TM}}{\upsilon_{TM}} \right)L } \frac{|g|^2}{2 \pi} \left| \frac{e^{\kappa L} - e^{-\kappa L}}{2 \kappa} \right|^2 ,  
\label{qnu5}
\end{equation}
\begin{equation}
    \left( \begin{array}{c}
Q^{\rm noise}_{TE\nu }(L) \\
Q^{\rm noise}_{TM(-\nu) }(L)
\end{array} \right) = 
\left( \begin{array}{c}
\frac{\Gamma_{TM}}{\upsilon_{TM}} \\
\frac{\Gamma_{TE}}{\upsilon_{TE}}
\end{array} \right) \frac{|g|^2}{4\pi |\kappa|^2} F\left(\mu_{\pm},L \right),
\label{qnoise}
\end{equation}
where 
\begin{equation}
F\left(\mu_{\pm},L \right) = \frac{e^{2{\rm Re}[\mu_{+}]L} - 1}{2 {\rm Re}[\mu_{+}]} + \frac{e^{2{\rm Re}[\mu_{-}]L} - 1}{2 {\rm Re}[\mu_{-}]} - 2  {\rm Re} \left[ \frac{e^{(\mu_{+}^* + \mu_{-})L} - 1}{\mu_{+}^* +\mu_{-}}  \right] ;
\label{Theta}
\end{equation}
\begin{eqnarray}
\mu_{\pm} &=& i\frac{\nu }{2}\left( \frac{1}{\upsilon_{TM}}+\frac{1}{\upsilon_{TE}}\right)  - \frac{1 }{2}\left( \frac{\Gamma_{TM}}{\upsilon_{TM}}+\frac{\Gamma_{TE}}{\upsilon_{TE}}\right) \pm \kappa, \label{mupm} \\
\kappa &=& \sqrt{\left\vert g\right\vert ^{2}- \frac{1}{4}\left[ D(\nu) + i \left( 
\frac{\Gamma_{TE}}{\upsilon_{TE}}- \frac{\Gamma_{TM}}{\upsilon_{TM}} \right) \right]^2   } , \label{kapa} \\
D(\nu) &=&  \nu \left( 
\frac{1}{\upsilon_{TE}}- \frac{1}{\upsilon_{TM}} \right). 
\label{dnu} 
\end{eqnarray}%
Here $D(\nu)$ is the phase mismatch for TE and TM modes at frequencies $\frac{\omega_p}{2} + \nu$ and $\frac{\omega_p}{2} - \nu$, respectively. 
When calculating the noise components of the fluxes we assumed that at optical frequencies the reservoir can be treated as having zero temperature. In this case we have $n_T(\omega) = 0$ in Eq.~(\ref{cf for L}). 

Note that the dynamic components of the fluxes in TE and TM modes are equal to each other even though their absorption losses may be very different; see Eq.~(\ref{qnu5}). This property holds only for vacuum boundary conditions with zero average number of photons. For a classical field or any multiquantum field at the boundary the mode with lower losses will accumulate a higher flux. 

The frequency spectrum of the downconverted photons is determined by the dependence $\kappa(\nu)$ in Eqs.~(\ref{kapa}), (\ref{dnu}).  
As follows from Eqs.~(\ref{kapa}), (\ref{dnu}), and (\ref{qnu5}),  in the absence of dissipation the parametric amplification occurs in the frequency interval
$|D(\nu)| = \left| \nu \left( 
\frac{1}{\upsilon_{TE}}- \frac{1}{\upsilon_{TM}} \right) \right| < 2 |g|$. For $D(\nu) \rightarrow 0$ the threshold for parametric amplification is determined by dissipation:  $ \frac{\Gamma_{TE} \Gamma_{TM}}{\upsilon_{TE}\upsilon_{TM}} < |g|^2$. Taking into account Eq.~(\ref{g}), the last inequality can be written as $\Gamma_{TE} \Gamma_{TM} < \frac{|A|^2}{\hbar^2}$, which is exactly the condition for the parametric decay in the initial-value problem \cite{tokman2019}. 

As one can see from Eq.~(\ref{qnoise}), the decay photon fluxes ``swap'' their noise components in the SPDC process: the photon flux in the TE mode is proportional to the absorption coefficient of the TM mode and vice versa. Therefore, when the noise reservoir is at zero temperature, there occurs parametric transfer of quantum noise between the two decay modes while  the photon flux of a given mode does not have any contribution from its own noise component. This feature is characteristic of the down-conversion and it illustrates that the contribution of noise always has to be included in the analysis as it is present even at zero temperature of the reservoir. In contrast,  one can show that in the {\it up-conversion} process the Langevin noise does not make any contribution to the upconverted photon flux as long as the reservoir can be treated as having zero temperature for high enough frequencies. 

It follows from Eqs.~(\ref{qnu5})-(\ref{Theta}) that the relative contribution of the Langevin noises is negligible in the parametric amplification regime when $|g| \gg \frac{\Gamma_{TE}}{\upsilon_{TE}}, \frac{\Gamma_{TM}}{\upsilon_{TM}} $. Although this limit is unrealistic for monolithic laser devices, we will still give the result for the spectral flux: 
\begin{equation}
Q_{TE\nu}(L) = Q_{TM(-\nu)}(L) \approx \frac{|g|^2 e^{-\left( 
\frac{\Gamma_{TE}}{\upsilon_{TE}} + \frac{\Gamma_{TM}}{\upsilon_{TM}} \right)L } }{2\pi |\kappa|^2} \left\{ 
\begin{array}{c} 
\sinh^2(|\kappa| L) \; {\rm for} \; D(\nu) < 2 |g| \\
\sin^2(|\kappa| L) \; {\rm for} \; D(\nu) > 2 |g|
\end{array}
\right. , 
\label{qnu33}
\end{equation}
where $|\kappa|^2 \approx \left| |g|^2 - \frac{1}{4} D^2(\nu) \right|$. Clearly, the flux of downconverted photons is nonzero even outside the parametric amplification bandwidth; however, it decays at large detunings as $\frac{1}{|\kappa|^2}$ and gets absorbed at propagation distances larger than the absorption length.

If the parametric gain is low, $|g| \ll \frac{\Gamma_{TE}}{\upsilon_{TE}}, \frac{\Gamma_{TM}}{\upsilon_{TM}} $, the flux of downconverted photons decays over the distances larger than the absorption length at all frequencies. This is the only realistic situation for a laser device, as one can see from the numerical estimates below. The expression for the flux is especially simple for propagation distances shorter than the absorption length, where the noise contribution is insignificant and we obtain  
\begin{equation}
Q_{TE\nu}(L) = Q_{TM(-\nu)}(L) = \frac{|g|^2 \sin^2(|\kappa| L)}{2 \pi |\kappa|^2}  , 
\label{qnu44}
\end{equation}
where $|\kappa| \approx \frac{1}{2} |D(\nu)|$.  These expressions for spectral flux densities have to be integrated over the bandwidth $\Delta \omega$ determined by the detection system to obtain the total flux. The result is in the Appendix, together with an alternative approach to obtain the 
nonperturbative solution to Eqs.~(\ref{ceq for cte}) and (\ref{ceq for ctm}) in the absence of dissipation following the Riemann-Volterra method.


\subsection{A numerical example for intracavity SPDC in the GaSb-based laser}

As a specific example, we calculate the performance of the proposed parametric source of biphotons using the device shown in Fig.~1. We consider the degenerate SPDC when the  pump wavelength 2032 nm and the wavelength of TE and TM-polarized decay photons is 4064 nm at exact phase matching, see the crossing point of phase matching curves in Fig.~1(d). For an intracavity pump power of 1 W and the waveguide width of 10 $\mu$m the coupling coefficient $g$ in Eq.~(\ref{g}) which according to Eq.~(\ref{kapa}) determines the maximum parametric gain is $ g = 0.08$ cm$^{-1}$. This number reflects the reduction by about a factor of 10 due to opposite symmetry of the pump and signal modes which leads to partial cancellation in the overlap integral in Eq.~(\ref{A}). 

Note that despite exact phase matching at central frequencies the waveguide dispersion leads to significant difference in the group velocities of the TE and TM decay modes: $\upsilon_{TE} \simeq 8.24 \times 10^9$ cm/s, whereas $\upsilon_{TM} \simeq 8.34 \times 10^9$ cm/s. This group velocity mismatch together with the magnitude of the parametric gain control the spectral properties of the generated biphotons.


\begin{figure}[htbp]
\centering\includegraphics[width=8cm]{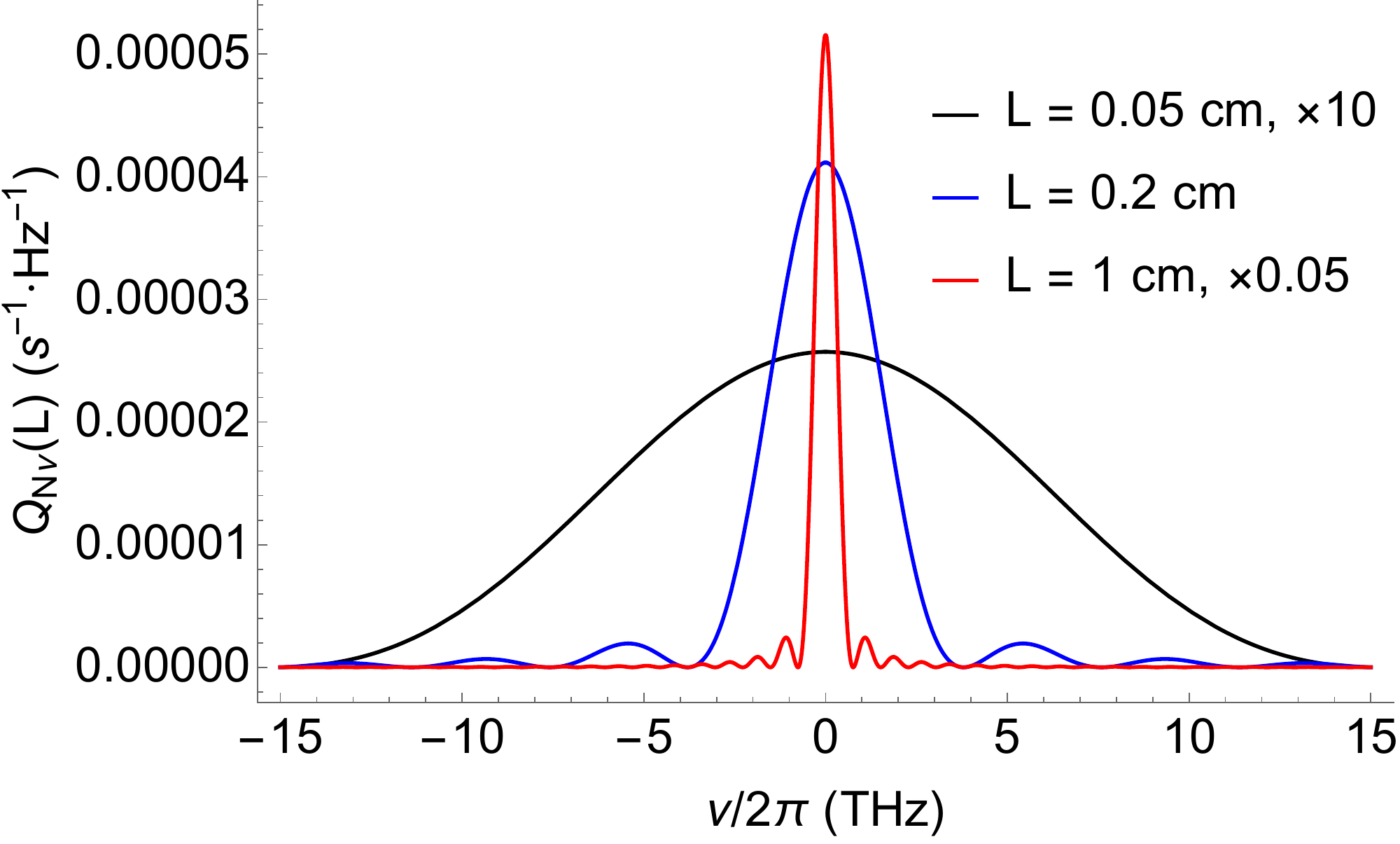}
\caption{Spectral flux density of the signal photons given by Eq.~(\ref{sqnu5}) for zero absorption losses at three different device lengths of 0.05 cm, 0.2 cm, and 1 cm. Here N = TE or TM. The horizontal axis is the frequency of detuning from resonance $\frac{\nu}{2 \pi}$ in THz. }
\label{fig:noloss}
\end{figure}

\begin{figure}[htbp]
\centering\includegraphics[width=8cm]{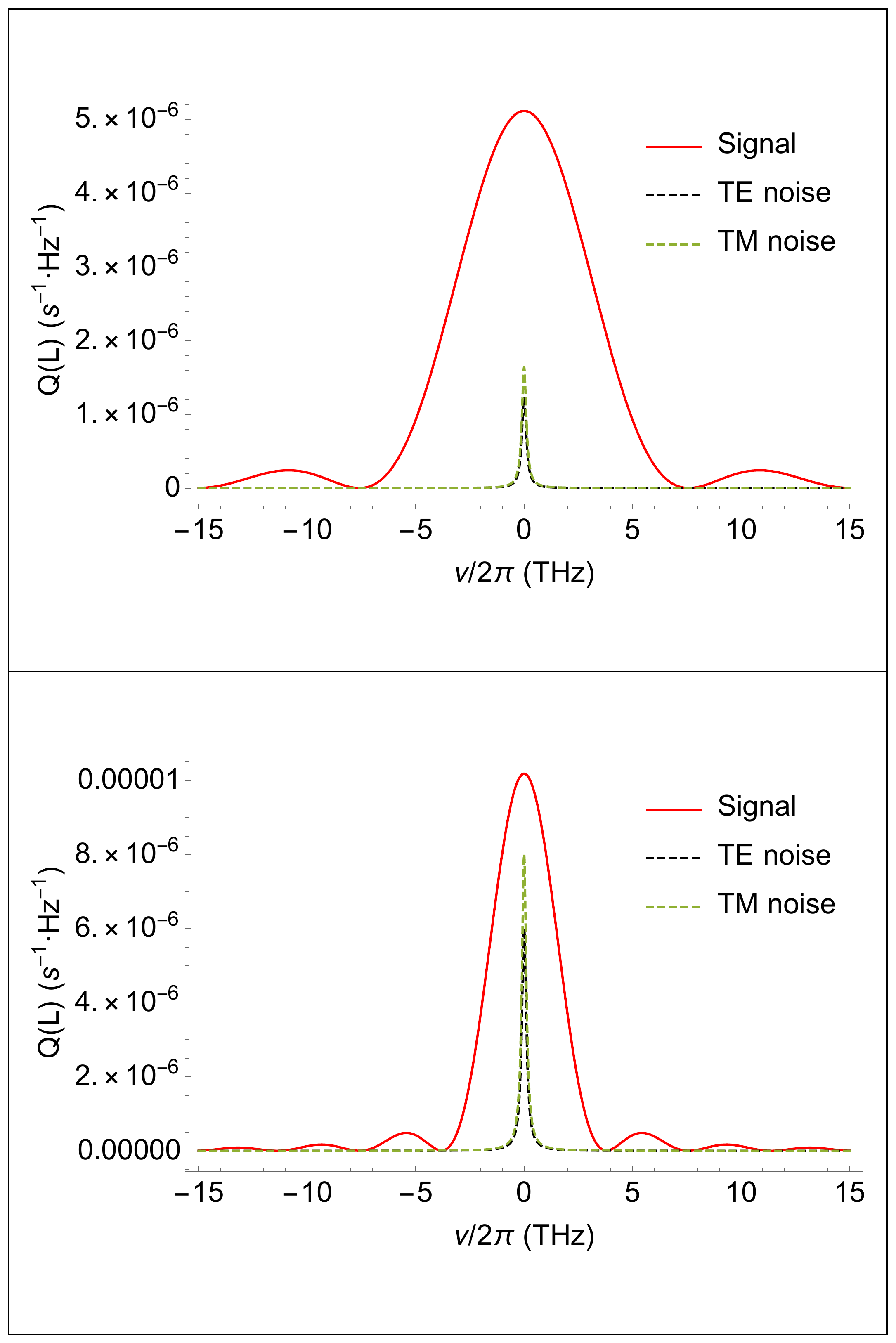}
\caption{Spectral flux density of the signal photons given by Eq.~(\ref{qnu5}) (solid curve) and TE- and TM-polarized noise photons given by Eq.~(\ref{qnoise}) (dashed curves) for the field absorption coefficients $ \frac{\Gamma_{TE}}{\upsilon_{TE}} = 4$ cm$^{-1}$ and  $\frac{\Gamma_{TM}}{\upsilon_{TM}} = 3$ cm$^{-1}$ at the device length of 1 mm (top panel) and 2 mm (bottom panel).  The horizontal axis is the frequency of detuning from resonance $\frac{\nu}{2 \pi}$ in THz. }
\label{fig:loss}

\end{figure}

For the sake of comparison, we start from the ideal case of negligible dissipation. Figure \ref{fig:noloss} shows the spectral fluxes of the parametric decay photons for different lengths of the device for negligible absorption of the field modes.  As follows from Eqs.~(\ref{skapa}), (\ref{sdnu}), and (\ref{sqnu5}),  the parametric amplification occurs in the relatively narrow frequency interval determined by $|\nu| \leq |g| \left( \frac{1}{\upsilon_{TE}}- \frac{1}{\upsilon_{TM}} \right)^{-1}$, or $\frac{\nu}{2\pi} \simeq 0.02$ THz for our parameters. This causes a sharp peak in the flux at low detunings for long enough propagation lengths $|g|L \geq 1$.  At much larger detunings $\kappa$ becomes  imaginary and scales as $\kappa \sim i \frac{|D(\nu)|}{2}$. In this case the signal flux scales according to Eq.~(\ref{sqnu44}) for low losses or short propagation lengths. Therefore, the total SPDC bandwidth defined as the spectral width of its main maximum is determined by $|\kappa|L < \pi$, or  
\begin{equation}
\label{totalb}
|\nu| < \frac{\Delta \omega_{tot}}{2} = \frac{2 \pi}{L}  \left( \frac{1}{\upsilon_{TE}}- \frac{1}{\upsilon_{TM}} \right)^{-1},
\end{equation}
 explaining strong dependence on the propagation length in Figs.~\ref{fig:noloss} and \ref{fig:loss}.


Now we include realistic modal losses and associated noise.  Figure \ref{fig:loss} shows the spectral fluxes of the parametric decay photons for different lengths of the device and realistic absorption losses and noise for a laser device: field absorption coefficients $ \frac{\Gamma_{TE}}{\upsilon_{TE}} = 4$ cm$^{-1}$ and  $\frac{\Gamma_{TM}}{\upsilon_{TM}} = 3$ cm$^{-1}$ (the intensity absorption would be two times higher). The spectral width of its main maximum which determines the total SPDC bandwidth is given by $|\kappa|L < \pi$, or  $|\nu| < \frac{4 \pi}{L}  \left( \frac{1}{\upsilon_{TE}}- \frac{1}{\upsilon_{TM}} \right)^{-1}$. The signal flux is exponentially decreasing for propagation lengths longer than the absorption length, i.e., $\left( 
\frac{\Gamma_{TE}}{\upsilon_{TE}} + \frac{\Gamma_{TM}}{\upsilon_{TM}} \right)L \geq 1$. At the same time, the peak noise flux becomes stronger than the peak signal flux at those lengths. The noise bandwidth is narrower than the signal's. It is determined by the condition $\frac{\Gamma_{TE}}{\upsilon_{TE}}, \frac{\Gamma_{TM}}{\upsilon_{TM}} \sim \nu \left( \frac{1}{\upsilon_{TE}}- \frac{1}{\upsilon_{TM}} \right) $.  Therefore, the optimal device length that maximizes the SPDC flux while still avoiding noise throughout most of the SPDC bandwidth is of the order of 1 mm, which happens to be also the optimal length for high-performance GaSb-based diode lasers. The total SPDC bandwidth $\frac{\Delta \nu}{2 \pi}$ for these lengths is around 10 THz. As one can see from Fig.~\ref{fig:loss}, the signal flux for a 2 mm long device within the bandwidth of  $\frac{\Delta \omega}{ 2 \pi} = 2$ THz near the peak is around $ 10^8$ biphotons/s for realistic losses, which  makes it interesting for applications, especially for such a small monolithic device. The flux into the total SPDC bandwidth will be several times higher. The peak flux can be further increased by increasing the intracavity pump field intensity or modal overlap at the expense of a more complicated device design.  


\subsection{Fluctuations and correlations between fluxes of decay photons}

The above results shed light on the kind of quantum correlations (or
entanglement) that could be detected in the decay photon fluxes in the laser output. Suppose
that one can detect the photon fluxes with a given polarization (TE or TM) within spectral bands $\Delta \omega
_{+}$ and $\Delta \omega _{-}$ that are symmetrically located around the central frequency $
\frac{\omega _{p}}{2}$, i.e., they have their central frequencies at $\frac{\omega _{p}}{2} \pm \delta \omega_0$, and have the
frequency bandwidth equal to $\Delta \omega $. In other words, the photon fluxes are detected within the frequency intervals $ \delta \omega_0 - \frac{\Delta \omega}{2} \leq \omega - 
\frac{\omega _{p}}{2} \leq \delta \omega_0 + \frac{\Delta \omega}{2}$ and $ - \delta \omega_0 - \frac{\Delta \omega}{2} \leq \omega - 
\frac{\omega _{p}}{2} \leq - \delta \omega_0 + \frac{\Delta \omega}{2}$, respectively.  Note that Eqs.~(\ref{rhomb1}) and (\ref{rhomb2}) couple pairwise the following
operators of the spectral field harmonics: $\hat{a}_{TE\nu }$ with $\hat{a}%
_{TM\left( -\nu \right) }^{\dagger }$ and $\hat{a}_{TM\left( -\nu \right) }$
with $\hat{a}_{TE\nu }^{\dagger }$, where $\nu = \omega -  \frac{\omega _{p}}{2}$ is defined as the detuning from the central frequency, in the same way as in previous subsections. The fluxes
of $TE$ and $TM$ photons \textit{between} the spectral intervals $\Delta
\omega _{+}$ and $\Delta \omega _{-}$ will be correlated. At the same
time, there are no correlations between the photon fluxes with different polarizations
\textit{within} each bandwidth $\Delta \omega _{+}$ or $\Delta \omega _{-}$.

We can determine the degree of correlations between photon fluxes quantitatively by calculating their correlation function,
\begin{eqnarray}
 \mathcal{K}(\tau) &=& \left \langle \, \int\displaylimits_{\Delta \omega_{+}}   \hat{a}_{TE\nu }^{\dagger }\,  e^{i \nu (t+ \tau)} d\nu \int\displaylimits_{\Delta \omega_{+}}  \hat{a}_{TE\nu } \, e^{-i \nu (t+ \tau)} d\nu 
 \int\displaylimits_{\Delta \omega_{-}}   \hat{a}_{TM\nu }^{\dagger } \, e^{i \nu t} d\nu \int\displaylimits_{\Delta \omega_{-}}  \hat{a}_{TM\nu } \, e^{-i \nu t} d\nu \right \rangle  \nonumber \\
 &  -  & \left \langle \, \int\displaylimits_{\Delta \omega_{+}}   \hat{a}_{TE\nu }^{\dagger } \, e^{i \nu } d\nu \int\displaylimits_{\Delta \omega_{+}}  \hat{a}_{TE\nu } \, e^{-i \nu t} d\nu \right \rangle \left \langle \, 
 \int\displaylimits_{\Delta \omega_{-}}   \hat{a}_{TM\nu }^{\dagger } \, e^{i \nu t} d\nu \int\displaylimits_{\Delta \omega_{-}}  \hat{a}_{TM\nu } \, e^{-i \nu t} d\nu \right \rangle,
 \label{corr1}
 \end{eqnarray}
 and comparing it with fluctuations of each flux, given by
 \begin{equation}
\mathcal{D}_{N} = \left \langle \left( \, \int\displaylimits_{\Delta \omega_{\pm}}   \hat{a}_{N\nu }^{\dagger }\,  e^{i \nu t} d\nu \int\displaylimits_{\Delta \omega_{\pm}}  \hat{a}_{N\nu } \, e^{-i \nu t} d\nu  \right)^2 \right \rangle  - \left \langle \,  \int\displaylimits_{\Delta \omega_{\pm}}   \hat{a}_{N\nu }^{\dagger }\,  e^{i \nu t} d\nu \int\displaylimits_{\Delta \omega_{\pm}}  \hat{a}_{N\nu } \, e^{-i \nu t} d\nu  \right \rangle^2.
 \label{fluct1}
 \end{equation}
 Here $N = TE,TM$ correspond to the top and bottom sign in $\Delta \omega_{\pm}$, respectively. 
 
 The dimensionless parameter characterizing the degree of correlations at the waveguide output $L$ is 
 \begin{equation} 
\Theta(L,\tau)  = \frac{\mathcal{K}(L,\tau)}{\sqrt{\mathcal{D}_{TE}(L) \mathcal{D}_{TM}(L)}}.
 \label{corr2} 
 \end{equation}
 It reaches the maximum value of 1 for completely correlated fluxes, and is smaller than 1 otherwise. The correlation time for the photon fluxes is just an inverse of the detection bandwidth, i.e. it is $\sim 1/\Delta \omega_{tot}$ if the photons are detected over the whole SPDC bandwidth in Eq.~(\ref{totalb}) and it is of the order of $1/\Delta \omega$ for a narrower bandwidth.  
 
 All terms on the right-hand side of Eq.~(\ref{corr2}) can be calculated from the solution for the flux operators $\hat{a}_{N\nu }(L)$ obtained in the Appendix. For an optimal case, we choose the frequency intervals $\Delta \omega_{\pm}$ outside the Langevin noise band in Fig.~3, when $D(\nu) \gg \frac{\Gamma_{TE}}{\upsilon_{TE}}, \frac{\Gamma_{TM}}{\upsilon_{TM}} $ and we can neglect the terms dependent on the the Langevin operators in the expressions for $\hat{a}_{N\nu }$. Using the equality $\langle 0 \vert  \hat{a}_{N\nu }^{\dagger }(0) \hat{a}_{N\nu' }(0) \hat{a}_{N\nu'' }^{\dagger }(0)  \hat{a}_{N\nu'''}(0) \vert 0 \rangle = 0$ and the commutation relation Eq.~(\ref{scr}) one can obtain 
 \begin{eqnarray}
\mathcal{D}_{TE}(L) &=& Q^{(s)}_{TE}(L) e^{-\left( 
\frac{\Gamma_{TE}}{\upsilon_{TE}} + \frac{\Gamma_{TM}}{\upsilon_{TM}} \right)L } \frac{|g|^2}{2 \pi} \int_{\Delta \omega} \, d\nu  \left| \frac{e^{\kappa L} K_{-} - e^{-\kappa L} K_{+} }{2 \kappa} \right|^2 ,  
\label{fluct2} \\
\mathcal{D}_{TM}(L) &=& Q^{(s)}_{TM}(L) e^{-\left( 
\frac{\Gamma_{TE}}{\upsilon_{TE}} + \frac{\Gamma_{TM}}{\upsilon_{TM}} \right)L } \frac{|g|^2}{2 \pi} \int_{\Delta \omega} \, d\nu  \left| \frac{e^{\kappa L} K_{+} - e^{-\kappa L} K_{-} }{2 \kappa} \right|^2 ,  
\label{fluct3}
\end{eqnarray}
and 
\begin{equation}
\mathcal{K}(L) =  e^{- 2 \left( 
\frac{\Gamma_{TE}}{\upsilon_{TE}} + \frac{\Gamma_{TM}}{\upsilon_{TM}} \right)L } \frac{|g|^4}{4 \pi^2}  \left \vert \int_{\Delta \omega} \, d\nu \, e^{-i \nu \tau}    \frac{\left( e^{\kappa L} K_{-} - e^{-\kappa L} K_{+} \right) \left(e^{\kappa^* L} - e^{-\kappa^* L} \right)  }{|2 \kappa|^2} \right \vert^2,
\label{corr4}
\end{equation}
where the functions $\kappa(\nu)$ and $K_{\pm}(\nu)$ are given by Eqs.~(\ref{skapa})-(\ref{sK}) whereas the values of fluxes $Q^{(s)}_{TE,TM}(L)$ are determined by integrating the flux spectral densities in Eq.~(\ref{qnu5}) over the spectral bandwidth. 

 In our example of a dissipative laser waveguide $|g| \ll  \frac{\Gamma_{TE}}{\upsilon_{TE}}, \frac{\Gamma_{TM}}{\upsilon_{TM}} $, in which case $ \kappa \simeq i |\kappa|$ and $K_{\pm} \simeq \frac{-D(\nu)}{2 g} \pm \frac{|\kappa|}{g}$ where one can without loss of generality assume that $g$ is real. This leads to further simplification of the above integrals, in which 
 $$
 e^{\kappa L} K_{-} - e^{-\kappa L} K_{+} \simeq -i \frac{D(\nu)}{g} \sin(|\kappa| L) - \frac{2 |\kappa| }{g} \cos(|\kappa| L),
 $$ 
 $$
 e^{\kappa L} K_{+} - e^{-\kappa L} K_{-} \simeq -i \frac{D(\nu)}{g} \sin(|\kappa| L) + \frac{2 |\kappa| }{g} \cos(|\kappa| L).
 $$
 
\begin{figure}[htbp]
\centering\includegraphics[width=8cm]{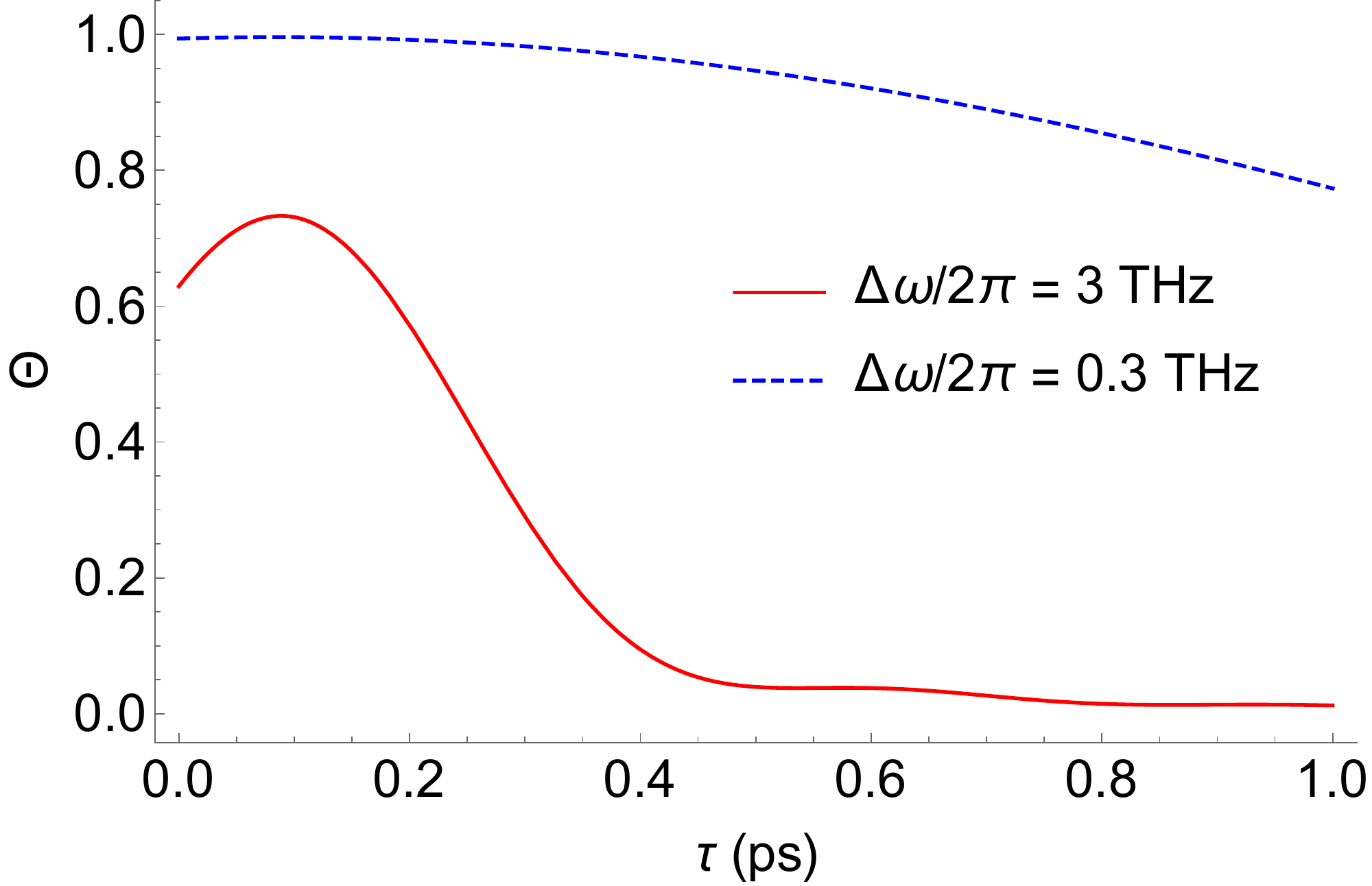}
\caption{The correlation parameter $\Theta(L,\tau)$ as a function of time delay $\tau$ for waveguide length  $L = 1$ mm and two values of signal bandwidths: $\frac{\Delta \omega}{2 \pi} = 3$ THz (red solid line) and $\frac{\Delta \omega}{2 \pi} = 0.3$ THz (blue dashed line). The frequency detuning of the signal bandwidth center from the central frequency $\frac{\omega_p}{2}$ is $\frac{\delta \omega_0}{2 \pi} = 6$ THz.  }
\label{fig:corr}

\end{figure}

 It is straightforward to calculate that if the signal bandwidths are selected narrow enough as compared to the total SPDC bandwidth defined in Eq.~(\ref{totalb}), namely 
 \begin{equation}
 \label{corrband}
\Delta \omega \ll \Delta \omega_{tot} = \frac{4 \pi}{L} \left( \frac{1}{\upsilon_{TE}} - \frac{1}{\upsilon_{TM}} \right)^{-1},
 \end{equation}
the photon fluxes at the output facet of a waveguide $z = L$ have a maximum possible correlation: $\Theta(L,\tau = 0)  = 1$. With increasing signal bandwidth the maximum correlation is reduced below 1 and its peak is shifted towards nonzero time delays. This behavior is illustrated in Fig.~\ref{fig:corr}. 


Fig.~\ref{fig:corr} shows the correlation parameter $\Theta(L,\tau)$ as a function of time delay $\tau$ for waveguide length  $L = 1$ mm when the  total SPDC bandwidth in Eq.~(\ref{totalb}) is $\frac{\Delta \omega_{tot}}{2 \pi}  \simeq 15$ THz, which corresponds to the average flux spectra shown in Fig.~\ref{fig:loss} (upper panel). We took the center frequencies of the decay photon bandwidths shifted by $\frac{\delta \omega_0}{2 \pi} = \pm 6$ THz from the central frequency $\frac{\omega_p}{2}$. Clearly, for a narrow signal bandwidth $\frac{\Delta \omega}{2 \pi} = 0.3$ THz the correlation is close to its maximum value of 1 over time delays shorter than $\sim 1/\Delta \omega$. When the signal bandwidth becomes comparable in magnitude to the total SPDC bandwidth, the correlations degrade.


\section{ Boundary-value problem for the Schr\"{o}dinger equation}

In the boundary-value problem solved in the previous section, the observables are determined using a constant Heisenberg-picture state vector $\left\vert \Psi
\left( t = 0\right) \right\rangle $ at the boundary $ z = 0$. Within the same approximation one can also
introduce the notion of space-dependent state vector which would be
equivalent to the space evolution of Heisenberg operators. With this goal in
mind, let's look at the equations (\ref{rhomb1}) and (\ref{rhomb2}) for the spectral components of the field
operators neglecting for simplicity the Langevin noise and dissipation. Since these equations contain only spatial derivatives, taking into account the commutation relations (\ref{cr}) for $\hat{a}_{N\nu }$ 
one can write the  ``lossless'' version of Eqs.~(\ref{rhomb1}) and (\ref{rhomb2}) and their Hermitian conjugates as ``spatial'' versions of the Heisenberg equations, namely 
\begin{equation}
\frac{\partial }{\partial z}\hat{O}=\frac{i}{\hbar }\left[ \hat{H}_{eff},%
\hat{O}\right] ,  \label{Heisenberg equation}
\end{equation}%
where $\hat{O}=\hat{a}_{TE\nu },\hat{a}_{TE\nu }^{\dagger },\hat{a}%
_{TM\left( -\nu \right) },\hat{a}_{TM\left( -\nu \right) }^{\dagger }$ and 
\begin{eqnarray}
\hat{H}_{eff} = 2\pi \hbar \left\{ \int_{\Delta \omega} d\nu \frac{\nu }{\upsilon_{TE}} 
\hat{a}_{TE\nu }^{\dagger }\hat{a}_{TE\nu } + \int_{\Delta \omega} d\nu \frac{\nu}{\upsilon_{TM}}
\hat{a}_{TM\nu}^{\dagger }\hat{a}_{TM\nu} -  \right. \nonumber \\ \left. 
\int  \int_{\Delta \omega \Delta \omega }d\nu d\nu
^{\prime} \delta\left(\nu +\nu^{\prime} \right) 
 \left( g  \hat{a}_{TE\nu }^{\dagger }\hat{a}_{TM\nu ^{\prime }}^{\dagger }+h.c.\right) 
  \right\}.
 \label{heff}
\end{eqnarray}
Note that the operator $\hat{H}_{eff} $ in Eq.~(\ref{Heisenberg equation}) generates translations along $z$, not time, and therefore it has the dimension of momentum.

The formal solution to Eq.~(\ref{Heisenberg equation}) has a standard form,
\begin{equation*}
\hat{O}\left( z\right) =e^{\frac{i}{\hbar }\hat{H}_{eff}z}\hat{O}\left(
0\right) e^{-\frac{i}{\hbar }\hat{H}_{eff}z}.
\end{equation*}%
Note that one can represent in this way the $z$-dependence for any
combination of operators, $\hat{O}\left( z\right) \Rightarrow \left( \hat{a}%
_{TE\nu }\right) ^{n}$ $\left( \hat{a}_{TE\nu }^{\dagger }\right) ^{l}\left(
\hat{a}_{TM\left( -\nu \right) }\right) ^{p}\left( \hat{a}_{TM\left( -\nu
\right) }^{\dagger }\right) ^{s}$. After requesting that the following
condition be met, $\left\langle \Psi \left( 0\right) \right\vert \hat{O}%
\left( z\right) \left\vert \Psi \left( 0\right) \right\rangle =\left\langle
\Psi \left( z\right) \right\vert \hat{O}\left( 0\right) \left\vert \Psi
\left( z\right) \right\rangle $ , we arrive at%
\begin{equation*}
\left\vert \Psi \left( z\right) \right\rangle =e^{-\frac{i}{\hbar }\hat{H}%
_{eff}z}\left\vert \Psi \left( 0\right) \right\rangle ,
\end{equation*}%
which gives the space evolution equation for the state vector,
\begin{equation}
i\hbar \frac{\partial }{\partial z}\left\vert \Psi \right\rangle =\hat{H}%
_{eff}\left\vert \Psi \right\rangle  \label{space evolution equation}
\end{equation}

For the classical pumping field there is no real need in using Eq.~(\ref%
{space evolution equation}) because the Heisenberg equations (\ref{rhomb1}) and (\ref{rhomb2}) are linear and can be easily
solved. The situation is different when the pumping field is quantized too,
for example if it is given by 
\begin{equation}
\mathbf{\hat{E}}_{p}\mathbf{=}\hat{c}_{p}\left( z,t\right) \mathbf{E}%
_{p}\left( \mathbf{r}_{\perp }\right) e^{ik_{p}\left( \omega _{p}\right)
z-i\omega _{p}t}+\hat{c}_{p}^{\dagger }\left( z,t\right) \mathbf{E}%
_{p}^{\ast }\left( \mathbf{r}_{\perp }\right) e^{-ik_{p}\left( \omega
_{p}\right) z+i\omega _{p}t},  \label{Ed}
\end{equation}%
where $\mathbf{E}_{p}\left( \mathbf{r}_{\perp }\right) $ is the
normalization amplitude given by Eq.~(\ref{normalization of the field}) where one should replace subscript  $N$ with $p$.

 Instead of Eqs.~(\ref{rhomb1}) and (%
\ref{rhomb2}) we now obtain%
\begin{equation}
-i\frac{\nu }{\upsilon_{TE}}\hat{a}_{TE\nu }+\frac{\partial }{\partial z}\hat{a}%
_{TE\nu }=iG\int \int_{\Delta \omega \Delta \omega }d\nu ^{\prime }d\nu ^{\prime \prime
}\delta \left( \nu +\nu ^{\prime }-\nu ^{\prime \prime }\right) \hat{a}%
_{p\nu ^{\prime \prime }}\hat{a}_{TM\nu ^{\prime }}^{\dagger },
\label{coupled eqs 1}
\end{equation}%
\begin{equation}
i\frac{\nu ^{\prime }}{\upsilon_{TM}}\hat{a}_{TM\nu ^{\prime }}^{\dagger }+\frac{%
\partial }{\partial z}\hat{a}_{TM\nu ^{\prime }}^{\dagger }=-iG^{\ast }\int
\int_{\Delta \omega \Delta \omega  }d\nu d\nu ^{\prime \prime }\delta \left( \nu +\nu
^{\prime }-\nu ^{\prime \prime }\right) \hat{a}_{p\nu ^{\prime \prime
}}^{\dagger }\hat{a}_{TE\nu },  \label{coupled eqs 2}
\end{equation}%
\begin{equation}
-i\frac{\nu ^{\prime \prime }}{\upsilon_{p}}\hat{a}_{p\nu ^{\prime \prime }}+\frac{%
\partial }{\partial z}\hat{a}_{p\nu ^{\prime \prime }}=iG^{\ast }\int
\int_{\Delta \omega \Delta \omega }d\nu d\nu ^{\prime }\delta \left( \nu +\nu ^{\prime
}-\nu ^{\prime \prime }\right) \hat{a}_{TE\nu }\hat{a}_{TM\nu ^{\prime }},
\label{coupled eqs 3}
\end{equation}%
where $\hat{a}_{p\nu ^{\prime \prime }}=\sqrt{\upsilon_{p}}\hat{c}_{p\nu ^{\prime
\prime }}$, $\upsilon_{p}$ is the group velocity of the pump mode, and
\begin{equation}
G=\frac{\int_{S}\mathbf{E}_{TE}^{\ast }\left( \mathbf{r}_{\perp }\right) %
\left[ \overleftrightarrow{\overleftrightarrow{\chi }}^{\left( 2\right)
}\left( \mathbf{r}_{\perp }\right) \mathbf{E}_{p}\left( \mathbf{r}_{\perp
}\right) \mathbf{E}_{TM}^{\ast }\left( \mathbf{r}_{\perp }\right) \right]
d^{2}r}{\hbar \sqrt{\upsilon_{TE}\upsilon_{TM}\upsilon_{p}}} .  \label{G}
\end{equation}

Equations~(\ref{coupled eqs 1})-(\ref{coupled eqs 3}) correspond to the
Heisenberg-like equation~(\ref{Heisenberg equation}) with effective ``Hamiltonian'' 
\begin{eqnarray}
\hat{H}_{eff} = 2\pi \hbar \left[ \int_{\Delta \omega} d\nu \frac{\nu }{\upsilon_{TE}} 
\hat{a}_{TE\nu }^{\dagger }\hat{a}_{TE\nu } + \int_{\Delta \omega} d\nu \frac{\nu}{\upsilon_{TM}}
\hat{a}_{TM\nu}^{\dagger }\hat{a}_{TM\nu} +  \int_{\Delta \omega} d\nu \frac{\nu }{\upsilon_{p}} 
\hat{a}_{p\nu }^{\dagger }\hat{a}_{p\nu }   \right. \nonumber \\ \left. 
- \int \int \int_{\Delta \omega \Delta \omega \Delta \omega }d\nu d\nu' d\nu'' \delta\left(\nu +\nu^{\prime} - \nu'' \right) 
 \left( G \hat{a}_{p\nu'' } \hat{a}_{TE\nu }^{\dagger }\hat{a}_{TM\nu ^{\prime }}^{\dagger }+h.c.\right) 
  \right] ,  \label{Heff1}
\end{eqnarray}%
i.e., one can again arrive at the equation of the type of Eq.~(\ref{space
evolution equation}), but with the ``Hamiltonian'' (\ref{Heff1}). 
The difference however is that now the operator-valued equations~(\ref%
{coupled eqs 1})-(\ref{coupled eqs 3}) are nonlinear whereas Eq.~(\ref{space
evolution equation}) for the state vector is always linear. This is a
crucial advantage of the approach based on Eq.~(\ref{space evolution
equation}). It is important that Eq.~(\ref{space evolution equation}) can be generalized
for open systems with dissipation and fluctuation effects using the
stochastic equation for the state vector \cite{tokman2021,chen2019}, and the method
of quantum jumps \cite{scully1997,plenio1998}. Here we illustrate our approach with an example of an external flux of pump photons propagating in a passive waveguide. Obviously, an active lasing device considered in the previous sections  cannot produce a single-photon pump flux. 

To avoid cumbersome derivations, we will switch from the continuous spectrum to a discrete set of frequencies; see, e.g., Ch.~10 in \cite{mandel}. This approach requires renormalization of the operators $\hat{a}_{N\nu}$, where $N = TE,TM$, or $p$. The quantities $\left\langle  \hat{a}_{N\nu}^{\dagger } \hat{a}_{N\nu} \right\rangle$ are now the total fluxes of photons of a given polarization within a given spectral line, i.e., they have the dimension of sec$^{-1}$. This renormalization of the operators is easiest to illustrate with an example of the parametric decay of a quasimonochromatic pump mode at frequency $\omega_p$ with bandwidth $\Delta \omega \ll \omega_p$.  The spectrum of signal and idler photons is convenient to represent as a set of discrete spectral lines at frequencies $\frac{\omega_p}{2} + \nu$, where $\nu$ span a discrete set of values symmetric with respect to $\omega_p/2$ and each spectral line has the same width $\Delta \omega$. The renormalized operators satisfy the commutation relations that follow from Eq.~(\ref{cr}) (see also the Supplemental Material in \cite{tokman2013}),  
\begin{equation}
\left[ \hat{a}_{N\nu },\hat{a}_{N^{\prime }\nu ^{\prime }}^{\dagger }\right]
=  \frac{\Delta \omega}{2 \pi} \delta _{NN^{\prime }}\delta _{\nu \nu ^{\prime }},  \label{comrel}
\end{equation}
where for $N = p$ the only option is $\nu = 0$. 
Therefore, one can introduce standard states of the boson field
\begin{equation}
\sqrt{\frac{2 \pi}{\Delta \omega}}  \hat{a}_{N\nu }\left\vert n_{N\nu }\right\rangle =\sqrt{n_{N\nu }}\left\vert
\left( n-1\right) _{N\nu }\right\rangle ,\ \ \sqrt{\frac{2 \pi}{\Delta \omega}}  \hat{a}_{N\nu }^{\dagger
}\left\vert n_{N\nu }\right\rangle =\sqrt{\left( n+1\right) _{N\nu }}%
\left\vert \left( n+1\right) _{N\nu }\right\rangle . \label{newcom}
\end{equation}
The discrete version of the effective `` Hamiltonian'' to be used in Eq.~(\ref{space evolution equation})  is
\begin{equation}
\hat{H}_{eff} = \frac{2\pi \hbar}{\Delta \omega} \sum_{\nu} \left[  \frac{\nu }{\upsilon_{TE}} 
\hat{a}_{TE\nu }^{\dagger }\hat{a}_{TE\nu } + \frac{\nu}{\upsilon_{TM}}
\hat{a}_{TM\nu}^{\dagger }\hat{a}_{TM\nu} -
 \left( G \hat{a}_{p} \hat{a}_{TE\nu }^{\dagger }\hat{a}_{TM(-\nu)}^{\dagger }+h.c.\right) 
  \right] .  \label{Heff}
\end{equation}
It is easy to verify that substituting $\hat{H}_{eff}$ from Eq.~(\ref{Heff}) into Eq.~(\ref{Heisenberg equation}) and taking into account the commutation relations (\ref{comrel}) will give a correct ``discrete'' version of Eqs.~(\ref{coupled eqs 1})-(\ref{coupled eqs 3}), see Eqs.~(\ref{dis1})-(\ref{dis3}) in the Appendix. 

When the state vector is expressed in terms of these number states as $\Psi_{N\nu} = \sum_n C_{N\nu}^{(n)}(z) |n \rangle$, the quantities $\left|C_{N\nu}^{(n)}(z) \right|^2 $ have the meaning of the probability to detect the flux of photons $\left\langle  \hat{a}_{N\nu}^{\dagger } \hat{a}_{N\nu} \right\rangle = Q_{0}n$ at the cross section $z$, where $Q_{0} = \frac{\Delta \omega}{2 \pi}$. The quantity $\hbar \omega_N Q_{0}$ is the energy flux transported by a single photon with waveform of duration $\frac{2 \pi}{\Delta \omega}$. The bandwidth $\Delta \omega$ and the values of the amplitudes $C_{N\nu}^{(n)}(z=0)$ at the boundary are determined by the properties of the pump. 

Consider a parametric decay when the quantum state at the boundary is $\left\vert \Psi \left( 0\right) \right\rangle = \left\vert 1_{p}\right\rangle
\left\vert 0_{TE,TM }\right\rangle $, where $\left\vert 0_{TE,TM }\right\rangle$ is a vacuum state of the signal and idler photons at all frequencies. In this case the solution to Eq.~(\ref{space evolution equation}) must have the form
\begin{eqnarray}
\left\vert \Psi \right\rangle &=& C_{p}\left( z\right) \left\vert 1_{p}\right\rangle
\left\vert 0_{TE,TM }\right\rangle \nonumber \\
&+& \sum_{\nu}
C_{W\nu}\left( z\right) \left\vert 0_{p}\right\rangle \left\vert 1_{TE\nu
}\right\rangle \left\vert 1_{TM(-\nu) }\right\rangle \prod_{\nu' \neq \nu, \nu'' \neq -\nu}
\left\vert 0_{TE\nu' }\right\rangle \left\vert 0_{TM\nu''}\right\rangle .  \label{sol1}
\end{eqnarray}%
All other states are forbidden by energy conservation. Equation (\ref{sol1}) is the generalization of a tripartite entangled state of the  Greenberger-Horne-Zeilinger (GHZ) type  \cite{tokman2021,dur,cunha,shalm,agusti}.


\begin{figure}[htbp]
\centering\includegraphics[width=8cm]{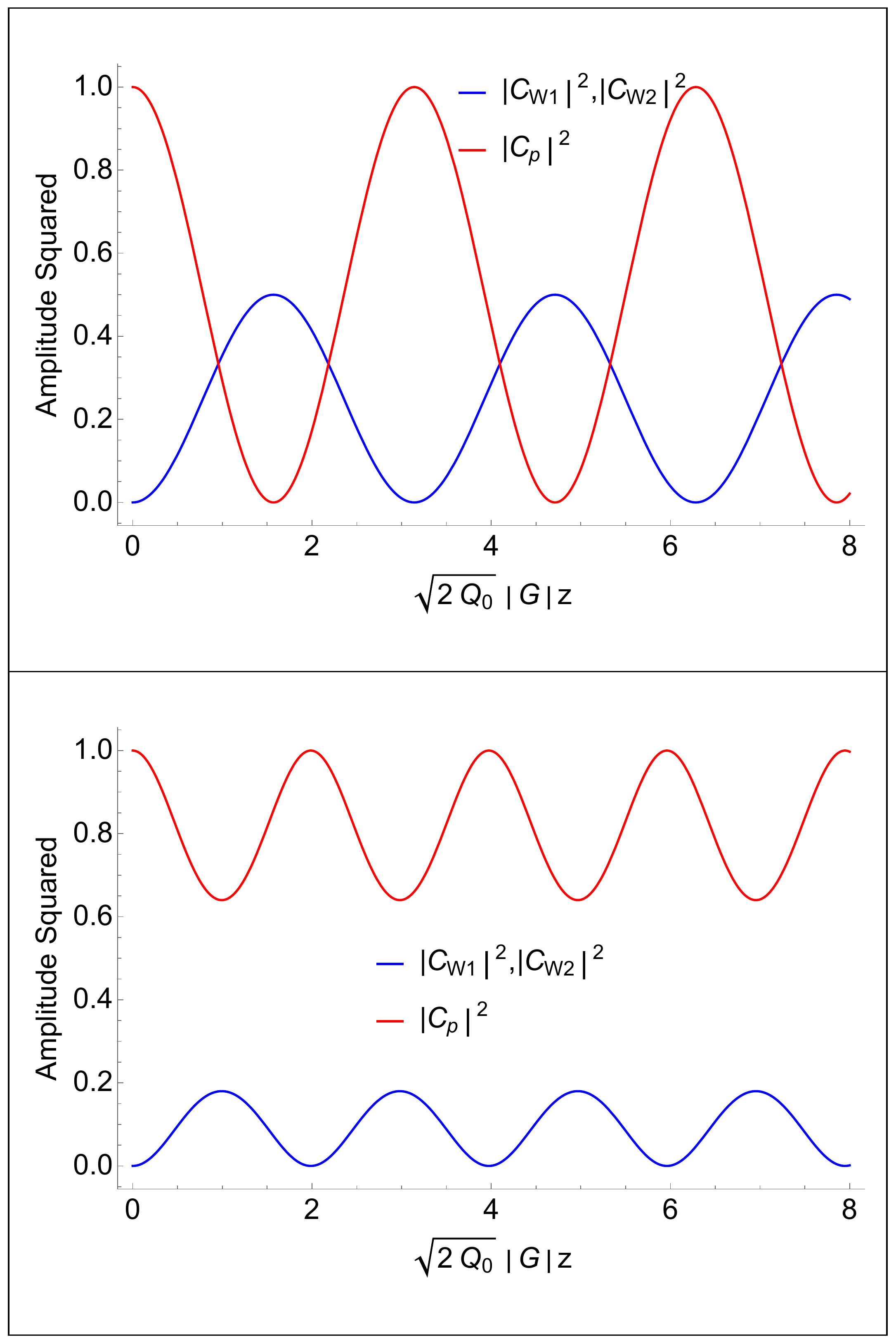}
\caption{Occupation probabilities $|C_{p}\left( z\right)|^2 $, $|C_{W1}\left( z\right)|^2
$ and $|C_{W2}\left( z\right)|^2 $  as a function of the normalized propagation distance $z$ along the waveguide  for $\delta = 0$ (top panel) and $\delta = 3 \sqrt{2 Q_{0} }|G| $ (bottom panel). The plots for $|C_{W1}\left( z\right)|^2
$ and $|C_{W2}\left( z\right)|^2 $ are identical. }
\label{fig:rabi}
\end{figure}


It is straightforward to solve coupled ordinary differential equations for the coefficients resulting from substituting Eq.~(\ref{sol1}) into Eq.~(\ref{space evolution equation}) with the ``Hamiltonian'' (\ref{Heff}). The detailed derivation is in the Appendix. Figure \ref{fig:rabi} illustrates the solution when the parametric decay of the pump occurs into photon pairs within only two symmetric spectral bands $\frac{\omega_p}{2} \pm \nu$, where $\nu$ has only one value. The figure shows $z$-dependence of the occupation probabilities $|C_{p}\left( z\right)|^2 $, $|C_{W1}\left( z\right)|^2
$ and $|C_{W2}\left( z\right)|^2 $ of the photon states $\left\vert 1_{p}\right\rangle
\left\vert 0_{TE,TM }\right\rangle $, $\left\vert 0_{p}\right\rangle \left\vert 1_{TE\nu
}\right\rangle \left\vert 0_{TM\nu }\right\rangle \left\vert 0_{TE\left(
-\nu \right) }\right\rangle \left\vert 1_{TM\left( -\nu \right)
}\right\rangle  $, and $\left\vert 0_{p}\right\rangle \left\vert 0_{TE\nu
}\right\rangle \left\vert 1_{TM\nu }\right\rangle \left\vert 1_{TE\left(
-\nu \right) }\right\rangle \left\vert 0_{TM\left( -\nu \right)
}\right\rangle $ respectively. The periodic modulation of the occupation probabilities with $z$ is a spatial analog of Rabi oscillations, in which the Rabi wavenumber $K_R$ for the probability amplitudes is given by 
\begin{equation}
K_{R}^{2}=\delta ^{2}+2 Q_{0} \left\vert G\right\vert^{2},
\label{Rabi wavenumber}
\end{equation}
where the detuning   
\begin{equation}
\delta =\nu \left( \frac{1}{\upsilon_{TM}} - \frac{1}{\upsilon_{TE}} \right).  \label{delta}
\end{equation}
For zero detuning from the central frequencies, at some values of $z$ there is a complete transfer of energy from the single-photon state of the pump to an entangled state of the decay photons. With increasing detuning the modulation occurs with a shorter spatial period according to Eq.~(\ref{Rabi wavenumber}) and the transfer of excitation is incomplete: it occurs with decreasing probability.  Note that the coefficients $C_{W1}\left( z\right)
$ and $C_{W2}\left( z\right) $ have the same amplitudes, although they may have different phases. 

By comparing the expression (\ref{g}) for $g$ in the previous section with the expression for the Rabi wavenumber $K_R = \sqrt{2 Q_{0}} |G|$ at zero detuning, one can verify that when the power of the classical pumping field in Eq. (\ref{g}) is equal to the power in the quantized single-photon flux $\hbar \omega_p \frac{\Delta \omega}{2 \pi}$, the expressions for $K_R$ and $g$ coincide, which is an important verification of the consistency of the two formalisms. 

Numerically, for the same waveguide design as in Fig.~1 one obtains $G \simeq 3 \times 10^{-11}$ s$^{1/2}$cm$^{-1}$. Assuming the bandwidth $\frac{\Delta \omega}{2 \pi} = 1$ THz the value of $K_R$ is very small, $K_R \sim 3 \times 10^{-5}$ cm$^{-1}$. This means that the probability of one incident pump photon to decay into the signal and idler photons after propagating the waveguide length of 3 cm is $10^{-4}$. Multiplying it by the flux of incident pump photons per second, one can obtain the expected flux of decay biphotons.

Here we considered the parametric decay into two relatively narrow spectral bands. For a broadband decay with the total width of the SPDC spectrum $\Delta \Omega \gg \Delta \omega$,  one can split the overall SPDC bandwidth into many narrow bands and perform the summation  over these bands in the resulting expression for $C_p$ using, e.g., the method developed in \cite{tokman2021-2} for strong coupling in the systems with inhomogeneous broadening of the spectra; see the Appendix for the derivation details. It is enough to calculate the expression for $C_p(z)$, since other amplitudes $C_{W\nu}(z)$ can be expressed through $C_p(z)$. Note that our results will not depend on the way we split the total bandwidth, i.e., on the parameter $\Delta \omega$.  

The general behavior of the solution for the probability amplitudes is controlled by the parameter 
$
    \alpha = \frac{\sqrt{\Delta \Omega}}{|G|}  \left|\frac{1}{\upsilon_{TM}} - \frac{1}{\upsilon_{TE}} \right|
$. 
When $\alpha \ll 1$, the solution for $C_p(z)$ is qualitatively similar to the one in the case of a parametric decay into two symmetric narrow lines as long as $\delta \ll \sqrt{2Q_{0}} |G|$; see Fig.~\ref{fig:rabi}.  All occupation probabilities $|C_{W\nu}(z)|^2$ are the same and they are related to $|C_{p}(z)|^2$ by conservation of the photon flux: $|C_{p}(z)|^2 + \sum_{\nu} |C_{W\nu}(z)|^2 = 1$. The Rabi wavenumber which describes spatial oscillations is given by 
\begin{equation}
K_{R} \approx |G| \sqrt{\frac{\Delta \Omega}{2\pi}}. 
\label{Rabi2}
\end{equation}
Since for two symmetric bands we have $\Delta \Omega = 2 \Delta \omega$, Eq.~(\ref{Rabi2}) coincides with Eq.~(\ref{Rabi wavenumber}) in the limit of small $\delta$. 

For the waveguide parameters shown in Fig.~1, we are in the regime corresponding to the opposite limit $\alpha \gg 1$. 
In this case the 
dephasing due to spectral broadening dominates and the probability amplitude  $C_p(z)$  decays exponentially along $z$ with an exponent 
\begin{equation}
    \kappa \simeq \frac{|G|^2}{2} \frac{1}{\left|\frac{1}{\upsilon_{TM}} - \frac{1}{\upsilon_{TE}} \right| }.
    \label{greater4}
\end{equation}
The attenuation rate $2\kappa$ of the occupation probability  due to spectral broadening is $\sim 10^{-9}$ cm$^{-1}$, indicating a very low rate of biphoton production, as expected for a single-photon pump.

\section{Conclusions}

In conclusion, we advanced the quantum theory of the SPDC of eigenmodes in finite-length semiconductor waveguides which takes into account not only all propagation effects such as phase and group velocity mismatch but also, most importantly, the effects of dissipation and quantum and thermal noise. The latter effects are crucial to include in the design of any monolithic semiconductor quantum device and in fact any emerging quantum photonic circuits that are based on lossy materials with high nonlinearity. For example, we show that for the SPDC process the quantum noise  makes significant and even dominant contribution within the signal/idler bandwidth even at low ambient temperature. Any experiment aimed at creating monolithic sources of quantum light has to take coupling to noisy reservoirs into account. Our paper provides the theory foundation and convenient analytic formulas to accomplish that. 

We applied our formalism to propose and evaluate the performance of a high-brightness, ultra-compact electrically pumped laser source of entangled photons generated by intracavity SPDC of lasing modes. The specific design in the paper is based on the III-Sb heterostructure and operation in the atmospheric transparency window of 3-5 $\mu$m wavelengths. However, the same device concept can be applied to any III-V material system at other wavelengths. 

We developed an approach based on the propagation equation for the state vector which solves the nonperturbative boundary-value problem of the parametric decay of a {\it quantized} single-photon pump mode and can include the effects of dissipation and noise. Our formalism is applicable to a wide variety of nonlinear wave mixing propagation problems in which all fields are quantized. It unifies the SPDC process with the strong coupling regime of cavity QED. The parametric strong coupling between three or more degrees of freedom leads to the formation of tripartite entangled states with many applications in quantum information and connections to other areas in quantum optics. 

\begin{acknowledgments}
The authors are grateful to Maria Erukhimova for helpful discussions. This work has been supported in part by National Science Foundation Awards No.~2135083 and 1936276, and Texas A\&M University through STRP, X-grant and T3-grant programs. M.T. acknowledges the support from the Russian Foundation for Basic Research Grant No. 20-02-00100. 
\end{acknowledgments}

\appendix

\section{Derivations for Sec.~IV of the main paper: the boundary-value problem for Heisenberg
operators}

Here we derive the equations that describe evolution of the operators $\hat{c}_{N}(z,t)$ which determine the
quantized field of decay photons. We use the mode index $N=TE,TM$ for brevity, which
labels both the field polarization and the transverse profile of the field $
\mathbf{E}_{N}\left( \mathbf{r}_{\perp }\right) $. Its dispersion equation
is $\omega =\omega _{N}\left( k\right) $. The normalization of the field is defined in Sec.~4 of the main paper; see Eq.~(15) there.  

The field operators for both quantized modes obey the wave equation%
\begin{equation}
\frac{\partial ^{2}}{\partial t^{2}}\left( \hat{\varepsilon}\mathbf{\hat{E}}%
\right) +c^{2}\nabla \times \nabla \times \mathbf{\hat{E}=-}4\pi \frac{%
\partial ^{2}}{\partial t^{2}}\delta \mathbf{\hat{P}}  \label{wave equation}
\end{equation}%
where%
\begin{equation*}
\hat{\varepsilon}\mathbf{\hat{E}=}\int_{0}^{\infty }\overleftrightarrow{%
\varepsilon }\left( \mathbf{r},\tau \right) \mathbf{\hat{E}}\left( \mathbf{r,%
}t-\tau \right) d\tau ,
\end{equation*}%
$\hat{\varepsilon}$ is a linear Hermitian operator; $\int_{0}^{\infty }%
\overleftrightarrow{\varepsilon }\left( \mathbf{r}_{\perp },\tau \right)
e^{i\omega \tau }d\tau =\overleftrightarrow{\varepsilon }\left( \omega ,%
\mathbf{r}_{\perp }\right) $ is the dielectric tensor for a nonuniform medium with frequency dispersion.
The operator 
\begin{equation}
\delta \mathbf{\hat{P}=}\delta \mathbf{\hat{P}}_{diss}+\delta \mathbf{\hat{P}%
}_{L}+\delta \mathbf{\hat{P}}_{nl}  \label{polarization operator}
\end{equation}%
includes the part describing linear dissipation (since we take as $\hat{%
\varepsilon}$ Hermitian), noise component of the polarization, and the
nonlinear polarization.

Within the slowly varying amplitude approximation, Eq.~(\ref{wave equation})
is reduced to%
\begin{equation}
\frac{\partial }{\partial t}\hat{c}_{N}+\upsilon_{N}\frac{\partial }{\partial z}%
\hat{c}_{N}+\Gamma _{N}\hat{c}_{N}=\hat{L}_{N}+\frac{i}{\hbar }e^{i\left[
\kappa _{N}-k_{N}\left( \omega _{N}\right) \right] z}\int_{S}\mathbf{E}%
_{N}^{\ast }\left( \mathbf{r}_{\perp }\right) \delta \mathbf{\hat{P}}%
_{nl;N}\left( \mathbf{r}_{\perp },t,z\right) d^{2}r.
\label{sEq for c operators}
\end{equation}%
Here%
\begin{equation}
\delta \mathbf{\hat{P}}_{nl}=\sum_{N=TE,TM}\left[ \delta \mathbf{\hat{P}}%
_{nl;N}\left( \mathbf{r}_{\perp },t,z\right) e^{i\kappa _{N}z-i\omega
_{N}t}+\delta \mathbf{\hat{P}}_{nl;N}^{\dagger }\left( \mathbf{r}_{\perp
},t,z\right) e^{-i\kappa _{N}z+i\omega _{N}t}\right] ,
\label{soperator of the nonlinear polarization}
\end{equation}%
where $\delta \mathbf{\hat{P}}_{nl;N}$ is the operator of the nonlinear
polarization at frequency $\omega _{N}$; $\Gamma _{N}$ determines modal
losses and is related to the Langevin noise operator $\hat{L}_{N}$ through
fluctuation-dissipation relations (see \cite{tokman20152,tokman2013,vdovin2013,tokman2016,erukhimova2017,tokman2018}). Following \cite{vdovin2013,tokman2016,erukhimova2017}, we
will use the following relationships for the Langevin noise operator,%
\begin{equation}
\left[ \hat{L}_{N\nu }\left( z\right) ,\hat{L}_{N^{\prime }\nu ^{\prime
}}^{\dagger }\left( z^{\prime }\right) \right] =\frac{\Gamma _{N}}{\pi }%
\delta _{NN^{\prime }}\delta \left( \nu -\nu ^{\prime }\right) \delta \left(
z-z^{\prime }\right) ,  \label{scommutator for L}
\end{equation}%
\begin{equation}
\left\langle \hat{L}^{\dagger}_{N\nu }\left( z\right) \hat{L}_{N^{\prime }\nu^{\prime
}}\left( z^{\prime }\right) \right\rangle =\frac{\Gamma
_{N}n_{T}\left( \omega _{N}\right) }{\pi }\delta _{NN^{\prime }}\delta
\left( \nu -\nu ^{\prime }\right) \delta \left( z-z^{\prime }\right) ,
\label{scf for L}
\end{equation}%
where $\left\langle \cdots \right\rangle $ means averaging over both an
initial quantum state in the Heisenberg picture and the statistics of the
dissipative reservoir, $n_{T}\left( \omega \right) =\left( e^{\hbar \omega
/T}-1\right) ^{-1}$,
\begin{equation*}
\hat{L}_{N}=\int_{\Delta \omega }\hat{L}_{N\nu }e^{-i\nu t}d\nu ,\ \ \hat{L}%
_{N}^{\dagger }=\int_{\Delta \omega }\hat{L}_{N\nu }^{\dagger }e^{i\nu
t}d\nu .
\end{equation*}%
Equation~(\ref{scommutator for L}) ensures the conservation of the commutation relations (Eq.~(\ref{commutation relations}) of the main paper) despite the presence of dissipation.

Besides energy conservation Eq.~(\ref{ec}) of the main paper one has to satisfy the momentum
conservation (phase matching) condition%
\begin{equation*}
\left\vert k_{p}\left( \omega _{p}\right) -k_{TE}\left( \omega _{TE}\right)
-k_{TM}\left( \omega _{TM}\right) \right\vert L\ll 1,
\end{equation*}%
where $L$ is the length of the SPDC region, in the simplest case the laser
waveguide length. For our device geometry
(see Fig.~1) the combination of energy and momentum conservation at a
given pump frequency $\omega _{p}$ can be satisfied for two pairs of
frequencies $\omega _{TE}$ and $\omega _{TM}$ (Fig.~1(d)). There
is also one value of frequency $\omega _{p}$ for which the frequencies of
decay photons become equal, $\omega_{TE}=\omega_{TM}=\omega_{p}/2$. We
will consider degenerate and non-degenerate SPDC separately.


\subsection{Non-degenerate case: space-time propagation problem, the
perturbation method}

Here we assume that the spectral widths of decay photons $\sim \Delta \omega
$ determined by phase-matching bandwidth are much lower than the distance
between their central frequencies: $\Delta \omega \ll \left\vert \omega
_{TE}-\omega _{TM}\right\vert $.

The resulting operator-valued equations for the two modes making up the
entangled two-photon state at the output are%
\begin{equation}
\frac{\partial }{\partial t}\hat{c}_{TE}+\upsilon_{TE}\frac{\partial }{\partial z}%
\hat{c}_{TE}+\Gamma _{TE}\hat{c}_{TE}=\hat{L}_{TE}+\frac{i}{\hbar }%
e^{-i\left( k_{TM}+k_{TE}-k_{p}\right) z}A\hat{c}_{TM}^{\dagger }
\label{seq for cte}
\end{equation}%
\begin{equation}
\frac{\partial }{\partial t}\hat{c}_{TM}^{\dagger }+\upsilon_{TM}\frac{\partial }{%
\partial z}\hat{c}_{TM}^{\dagger }+\Gamma _{TM}\hat{c}_{TM}^{\dagger }=\hat{L%
}_{TM}^{\dagger }-\frac{i}{\hbar }e^{i\left( k_{TM}+k_{TE}-k_{p}\right)
z}A^{\ast }\hat{c}_{TE}  \label{seq for ctm}
\end{equation}%
Here%
\begin{equation}
A=\int_{S}\mathbf{E}_{TE}^{\ast }\left( \mathbf{r}_{\perp }\right) \left[
\overleftrightarrow{\overleftrightarrow{\chi }}^{\left( 2\right) }\left(
\mathbf{r}_{\perp }\right) \mathbf{E}_{p}\left( \mathbf{r}_{\perp
}\right) \mathbf{E}_{TM}^{\ast }\left( \mathbf{r}_{\perp }\right) \right]
d^{2}r,  \label{sA}
\end{equation}%
where $\overleftrightarrow{\overleftrightarrow{\chi }}^{\left( 2\right) }$
is the second-order nonlinear susceptibility, and 
\begin{equation*}
\mathbf{E}_{TE}^{\ast }%
\overleftrightarrow{\overleftrightarrow{\chi }}_{p}^{\left( 2\right) }%
\mathbf{E}_{p}\mathbf{E}_{TM}^{\ast }=\mathbf{E}_{TM}^{\ast }%
\overleftrightarrow{\overleftrightarrow{\chi }}_{p}^{\left( 2\right) }%
\mathbf{E}_{p}\mathbf{E}_{TE}^{\ast }; 
\end{equation*}
see \cite{keldysh1994}.

Equations~(\ref{seq for cte})-(\ref{seq for ctm}) have the boundary conditions%
\begin{equation}
\hat{c}_{N}\left( t,z=0\right) =\hat{c}_{N}^{\left( 0\right) }\left(
t\right) .  \label{sboundary conditions}
\end{equation}%
The slow time dependence in $\hat{c}_{N}^{\left( 0\right) }$ is due to a
finite (although narrow) bandwidth $\Delta \omega $,%
\begin{equation}
\hat{c}_{N}^{\left( 0\right) }\left( t\right) =\int_{\Delta \omega }\hat{c}%
_{N\nu }^{\left( 0\right) }e^{-i\nu t}d\nu ,\ \ \hat{c}_{N}^{\left( 0\right)
\dagger }\left( t\right) =\int_{\Delta \omega }\hat{c}_{N\nu }^{\left(
0\right) \dagger }e^{i\nu t}d\nu ,  \label{sslow time dependence}
\end{equation}%
where $\hat{c}_{N\nu }^{\left( 0\right) }$ is the Schr\"{o}dinger (constant)
operator. If the field at the boundary is an incoherent noise field with a
certain spectral photon distribution $n\left( \omega \right) $ , the
following useful relationships are satisfied:%
\begin{equation}
\left\langle \hat{c}_{N\nu }^{\left( 0\right) \dagger }\hat{c}_{N^{\prime
}\nu ^{\prime }}^{\left( 0\right) }\right\rangle =n\left( \omega _{N}\right)
\delta _{NN^{\prime }}\frac{\delta \left( \nu -\nu ^{\prime }\right) }{2\pi
\upsilon_{N}},\ \ \ \left\langle \hat{c}_{N\nu }^{\left( 0\right) }\hat{c}%
_{N^{\prime }\nu ^{\prime }}^{\left( 0\right) \dagger }\right\rangle =\left[
n\left( \omega _{N}\right) +1\right] \delta _{NN^{\prime }}\frac{\delta
\left( \nu -\nu ^{\prime }\right) }{2\pi \upsilon_{N}}  \label{srelationships}
\end{equation}%
In particular, for vacuum boundary conditions in Eq.~(\ref{srelationships})
we have $n\left( \omega _{N}\right) =0$. The corresponding photon flux in
the narrow frequency band $\Delta \omega $ is $Q_{N}=\upsilon_{N}\left\langle \hat{c%
}_{N}^{\left( 0\right) \dagger }\hat{c}_{N}^{\left( 0\right) }\right\rangle
=n\left( \omega _{N}\right) \frac{\Delta \omega }{2\pi }$. For a thermal
noise we have $n_{T}\left( \omega_N \right) =\left( e^{\hbar \omega_N 
/T}-1\right) ^{-1}$, which is reduced to $Q_{N}\approx \frac{T\Delta \omega
}{2\pi \hbar \omega _{i}}$ in the Rayleigh-Jeans limit. The last expression
corresponds to the known result: the radiation power $\frac{T\Delta \omega }{%
2\pi }$ received by a matched antenna in the black-body bath does not depend
on the size and shape of an aperture.

To start with the simplest case, we assume that the length $L$ of the decay region is
smaller than all absorption lengths $\frac{\upsilon_{N}}{\Gamma _{N}}$. This allows us to neglect dissipative and Langevin terms, 
\begin{equation}
\frac{\partial }{\partial t}\hat{c}_{TE}+\upsilon_{TE}\frac{\partial }{\partial z}%
\hat{c}_{TE}=\frac{i}{\hbar }e^{-i\left( k_{TM}+k_{TE}-k_{p}\right) z}A\hat{c%
}_{TM}^{\dagger }  \label{sEq of cte}
\end{equation}%
\begin{equation}
\frac{\partial }{\partial t}\hat{c}_{TM}^{\dagger }+\upsilon_{TM}\frac{\partial }{%
\partial z}\hat{c}_{TM}^{\dagger }=-\frac{i}{\hbar }e^{i\left(
k_{TM}+k_{TE}-k_{p}\right) z}A^{\ast }\hat{c}_{TE}  \label{sEq of ctm}
\end{equation}
When treating the degenerate SPDC in the next subsection, we will consider arbitrary propagation lengths and fully include the effects of dissipation and noise.

The formal solutions to Eqs.~(\ref{sEq of cte}) and (\ref{sEq of ctm}) are%
\begin{equation}
\hat{c}_{TE}=\hat{c}_{TE}^{\left( 0\right) }\left( t-\frac{z}{\upsilon_{TE}}\right)
+\frac{i}{\hbar }\frac{A}{\upsilon_{TE}}\int_{0}^{z}e^{-i\left(
k_{TM}+k_{TE}-k_{p}\right) \zeta }\hat{c}_{TM}^{\dagger }\left( t-\frac{%
z-\zeta }{\upsilon_{TE}},\zeta \right) d\zeta  \label{sfs for cte}
\end{equation}%
\begin{equation}
\hat{c}_{TM}^{\dagger }=\hat{c}_{TM}^{\left( 0\right) \dagger }\left( t-%
\frac{z}{\upsilon_{TM}}\right) -\frac{i}{\hbar }\frac{A^{\ast }}{\upsilon_{TM}}%
\int_{0}^{z}e^{i\left( k_{TM}+k_{TE}-k_{p}\right) \zeta }\hat{c}_{TE}\left(
t-\frac{z-\zeta }{\upsilon_{TM}},\zeta \right) d\zeta  \label{sfs for ctm}
\end{equation}%
where%
\begin{equation*}
\hat{c}_{TE}\left( t-\frac{z-\zeta }{\upsilon_{TM}},\zeta \right) =\hat{c}%
_{TE}\left( t,z\right) _{\left\{ t\Rightarrow t-\frac{z-\zeta }{\upsilon_{TM}}\
z\Rightarrow \zeta \right\} },\ \hat{c}_{TM}^{\dagger }\left( t-\frac{%
z-\zeta }{\upsilon_{TE}},\zeta \right) =\hat{c}_{TM}^{\dagger }\left( t,z\right)
_{\left\{ t\Rightarrow t-\frac{z-\zeta }{\upsilon_{TE}}\ z\Rightarrow \zeta
\right\} }\ .
\end{equation*}%
\textit{Within the perturbation expansion in terms of the coupling parameter}
$A$ we substitute unperturbed operators given by the first terms in the
right-hand side of Eqs.~(\ref{sfs for cte})-~(\ref{sfs for ctm}) (which
describe the transfer of the boundary conditions with the group velocity)
into the integrands in Eqs.~(\ref{sfs for cte}),(\ref{sfs for ctm}), namely,
\begin{eqnarray*}
\hat{c}_{TE}\left( t-\frac{z-\zeta }{\upsilon_{TM}},\zeta \right) &\Rightarrow &%
\hat{c}_{TE}^{\left( 0\right) }\left( t-\frac{z}{\upsilon_{TM}}+\zeta \frac{%
\upsilon_{TM}-\upsilon_{TE}}{\upsilon_{TE}\upsilon_{TM}}\right) , \\
\ \hat{c}_{TM}^{\dagger }\left( t-\frac{z-\zeta }{\upsilon_{TE}},\zeta \right)
&\Rightarrow &\hat{c}_{TM}^{\left( 0\right) \dagger }\left( t-\frac{z}{\upsilon_{TE}%
}+\zeta \frac{\upsilon_{TM}-\upsilon_{TE}}{\upsilon_{TE}\upsilon_{TM}}\right) \ .
\end{eqnarray*}%
This gives%
\begin{equation}
\hat{c}_{TE}\left( t,z\right) =\hat{c}_{TE}^{\left( 0\right) }\left( t-\frac{%
z}{\upsilon_{TE}}\right) +\frac{i}{\hbar }\frac{A}{\upsilon_{TE}}\int_{0}^{z}e^{-i\left(
k_{TM}+k_{TE}-k_{p}\right) \zeta }\hat{c}_{TM}^{\left( 0\right) \dagger
}\left( t-\frac{z}{\upsilon_{TE}}+\zeta \frac{\upsilon_{TM}-\upsilon_{TE}}{\upsilon_{TE}\upsilon_{TM}}\right)
d\zeta ,  \label{scte operator}
\end{equation}%
\begin{equation}
\hat{c}_{TM}^{\dagger }\left( t,z\right) =\hat{c}_{TM}^{\left( 0\right)
\dagger }\left( t-\frac{z}{\upsilon_{TM}}\right) -\frac{i}{\hbar }\frac{A^{\ast }}{%
\upsilon_{TM}}\int_{0}^{z}e^{i\left( k_{TM}+k_{TE}-k_{p}\right) \zeta }\hat{c}%
_{TE}^{\left( 0\right) }\left( t-\frac{z}{\upsilon_{TM}}+\zeta \frac{\upsilon_{TM}-\upsilon_{TE}}{%
\upsilon_{TE}\upsilon_{TM}}\right) d\zeta .  \label{sctm operator}
\end{equation}%
Expressions~(\ref{scte operator})-(\ref{sctm operator}) allow us to calculate
any experimental observables. For example, we can calculate the photon
fluxes within the bandwidth $\Delta \omega $ in the cross section $z=L$ for
vacuum boundary conditions:%
\begin{equation}
Q_{TE}=\upsilon_{TE}\left\langle \hat{c}_{TE}^{\dagger }\left( L\right) \hat{c}%
_{TE}\left( L\right) \right\rangle \approx \frac{\Delta \omega }{2\pi }\frac{%
\left\vert A\right\vert ^{2}}{\hbar ^{2}\upsilon_{TE}\upsilon_{TM}}\left\vert
\int_{0}^{L}e^{i\left( k_{TM}+k_{TE}-k_{p}\right) z}dz\right\vert ^{2}
\label{sqte}
\end{equation}%
\begin{equation}
Q_{TM}=\upsilon_{TM}\left\langle \hat{c}_{TM}^{\dagger }\left( L\right) \hat{c}%
_{TM}\left( L\right) \right\rangle =Q_{TE}  \label{sqtm}
\end{equation}%
The last equality corresponds to Manley-Rowe relations \cite{bloembergen1996}. The expression (\ref{sqte}) is valid when the   bandwidth $\Delta \omega $ satisfies $L \Delta \omega \left| \frac{1}{\upsilon_{TE}} - \frac{1}{\upsilon_{TM}} \right| \ll 1$. 

Using the spectral decomposition of the field operators given by Eqs.~(17) of the main paper, one can obtain the solutions for the spectral amplitudes
beyond the perturbation approach. We will present such a solution for the
degenerate case below, because this is the most interesting case for most
applications.


\subsection{Degenerate case: the nonperturbative solution for spectral amplitudes}

Consider now the degenerate SPDC when $\omega _{TE}=\omega _{TM}=\omega
_{p}/2$. 
We start with the most general case when there is still finite phase mismatch $\delta k$ at central frequencies $\omega_{TE} = \omega_{TM} = \omega_p/2$, namely
$$ k_{TM}\left(\frac{\omega_p}{2} \right) + k_{TE}\left(\frac{\omega_p}{2} \right) -k_p(\omega_p) = \delta k,
$$
and the field dissipation and Langevin noises cannot be neglected.
The coupled
equations for the field operators are%
\begin{equation}
\left( \frac{\partial }{\partial t}+ \Gamma_{TE} +\upsilon_{TE}\frac{\partial }{\partial z} \right)
\hat{c}_{TE} - \frac{i}{\hbar }A\hat{c}_{TM}^{\dagger } e^{-i \delta k z} = \hat{L}_{TE},
\label{sceq for cte}
\end{equation}%
\begin{equation}
\left( \frac{\partial }{\partial t} + \Gamma_{TM} +\upsilon_{TM}\frac{\partial }{%
\partial z} \right) \hat{c}_{TM}^{\dagger } + \frac{i}{\hbar }A^{\ast }\hat{c}_{TE} e^{i \delta k z} = \hat{L}^{\dagger}_{TM}.
\label{sceq for ctm}
\end{equation}

In the boundary-value problem, it is convenient to transfer from the
operators $\hat{c}_{N }$ which
determine the \textit{ density} of the photon number per unit length
along the waveguide, $\left\langle \hat{c}_{N }^{\dagger }\hat{c}%
_{N }\right\rangle $, to the operators $\hat{a}_{N }=\sqrt{\upsilon_{N}}\hat{c%
}_{N }$  which determine the \textit{flux} of photons in
the waveguide, $\left\langle \hat{a}_{N }^{\dagger }\hat{a}%
_{N }\right\rangle $. The operators $\hat{a}_{N }$ satisfy the
commutation relations that follow from Eq.~(16) of the main paper,
namely%
\begin{equation}
\left[ \hat{a}_{N\nu}\left( z\right) ,\hat{a}_{N^{\prime }\nu
^{\prime }}^{\dagger }\left( z\right) \right] =\delta _{NN^{\prime }}%
\frac{\delta \left( \nu-\nu ^{\prime }\right) }{2\pi },
\label{scr}
\end{equation}
where 
\begin{equation}
\hat{a}_{N}\left( z,t\right) =\int_{\Delta \omega }\hat{a}%
_{N\nu }e^{-i\nu t}d\nu ,\ \ \hat{a}_{N}^{
\dagger }\left( z,t\right) =\int_{\Delta \omega }\hat{a}_{N\nu }^{ \dagger }e^{i\nu t}d\nu.   
\label{sanu}
\end{equation}%
Vacuum boundary conditions for the flux operators follow from Eqs.~(\ref{srelationships}):
\begin{equation}
\left\langle \hat{a}_{TE\nu }^{\left( 0\right) \dagger }\hat{a}_{TE\nu^{\prime }}^{\left( 0\right) }\right\rangle =
\left\langle \hat{a}_{TM(-\nu) }^{\left( 0\right) \dagger }\hat{a}_{TM(-\nu^{\prime })}^{\left( 0\right) }\right\rangle =0
,\ \ \ \left\langle \hat{a}_{TE\nu }^{\left( 0\right) }\hat{a}_{TE\nu^{\prime }}^{\left( 0\right)\dagger }\right\rangle =
\left\langle \hat{a}_{TM(-\nu) }^{\left( 0\right) }\hat{a}_{TM(-\nu^{\prime })}^{\left( 0\right)\dagger }\right\rangle = \frac{\delta
\left( \nu -\nu ^{\prime }\right) }{2\pi }  \label{sabound}
\end{equation}%
where $ \hat{a}_{N\nu }^{\left( 0\right) } = \hat{a}_{N\nu }(z = 0) $.

Next, we transfer to the flux operators in Eqs.~(\ref{sceq for cte}) and (\ref{sceq
for ctm}) and use the Fourier expansion (\ref{sanu}). To get rid of the explicit $z$-dependence in the left-hand sides of Eqs.~(\ref{sceq for cte}) and (\ref{sceq
for ctm}) we make the substitution $\hat{a}_{TM(-\nu) }^{ \dagger } = \hat{\tilde{a}}_{TM(-\nu) }^{ \dagger } e^{i\frac{\delta k}{2} z}$, $\hat{a}_{TE\nu }^{ } = \hat{\tilde{a}}_{TE\nu }^{  } e^{-i\frac{\delta k}{2} z}$. 
This results in
\begin{equation}
\left( -i\frac{\nu + i \Gamma_{TE} }{\upsilon_{TE}} - i \frac{\delta k}{2}  +\frac{\partial }{\partial z} \right) \hat{\tilde{a}}_{TE\nu } - ig\hat{\tilde{a}}_{TM\left( -\nu \right) }^{\dagger } =\frac{1}{\sqrt{\upsilon_{TE}}} \hat{L}_{TE\nu}(z)  e^{-i \frac{\delta k}{2} z}  ,
\label{srhomb1}
\end{equation}
\begin{equation}
\left( -i\frac{\nu + i \Gamma_{TM} }{\upsilon_{TM}} + i \frac{\delta k}{2} +\frac{%
\partial }{\partial z} \right) \hat{\tilde{a}}_{TM\left( -\nu \right) }^{\dagger } + ig^{\ast }%
\hat{\tilde{a}}_{TE\nu } = \frac{1}{\sqrt{\upsilon_{TM}}} \hat{L}^{\dagger}_{TM(-\nu)}(z)  e^{i \frac{\delta k}{2} z},  \label{srhomb2}
\end{equation}%
where the coupling coefficient 
\begin{equation}
g=\frac{A}{\hbar \sqrt{\upsilon_{TE}\upsilon_{TM}}}.  \label{sg}
\end{equation}%

The solution of Eqs.~(\ref{srhomb1}) and (%
\ref{srhomb2}) can be written as (see the similar derivations in \cite{tokman2013,vdovin2013})
\begin{eqnarray} 
\left( \begin{array}{c}
\hat{a}_{TE\nu }(z) \\
\hat{a}_{TM(-\nu) }^{\dagger }z 
\end{array} \right)
 &=& e^{\mu_{+} z}
\left( \begin{array}{c}
e^{-i \frac{\delta k}{2} z} \\
e^{i \frac{\delta k}{2} z} K_{+} 
\end{array} \right) 
\left( \hat{U}_{+} + \int_0^z e^{- \mu_{+} \xi} \hat{F}_{+}(\xi) d\xi \right) \nonumber \\
&+& e^{\mu_{-} z}
\left( \begin{array}{c}
e^{-i \frac{\delta k}{2} z} \\
e^{i \frac{\delta k}{2} z} K_{-} 
\end{array} \right) 
\left( \hat{U}_{-} + \int_0^z e^{- \mu_{-} \xi} \hat{F}_{-}(\xi) d\xi \right),
\label{ssolution for ceq}
\end{eqnarray}%
where
\begin{eqnarray}
\mu_{\pm} &=& i\frac{\nu }{2}\left( \frac{1}{\upsilon_{TM}}+\frac{1}{\upsilon_{TE}}\right)  - \frac{1 }{2}\left( \frac{\Gamma_{TM}}{\upsilon_{TM}}+\frac{\Gamma_{TE}}{\upsilon_{TE}}\right) \pm \kappa, \label{smupm} \\
\kappa &=& \sqrt{\left\vert g\right\vert ^{2}- \frac{1}{4}\left[ D(\nu) + i \left( 
\frac{\Gamma_{TE}}{\upsilon_{TE}}- \frac{\Gamma_{TM}}{\upsilon_{TM}} \right) \right]^2   } , \label{skapa} \\
D(\nu) &=& \delta k + \nu \left( 
\frac{1}{\upsilon_{TE}}- \frac{1}{\upsilon_{TM}} \right), 
\label{sdnu} \\
K_{\pm} &=&  \frac{ - D(\nu) - i \left( 
\frac{\Gamma_{TE}}{\upsilon_{TE}}- \frac{\Gamma_{TM}}{\upsilon_{TM}} \right) }{2g} 
\mp i \frac{\kappa }{g}  , 
\label{sK}  \\
\hat{U}_{\pm} &=&  \pm \frac{g}{i2\kappa } \left(\hat{a}_{TE\nu }(0) K_{\mp} - \hat{a}_{TM( -\nu) }^{ \dagger }(0) \right),
\label{sU} \\
 \hat{F}_{\pm}(\xi) &=& \pm \frac{g}{i2\kappa } \left( K_{\mp} \frac{1}{\sqrt{\upsilon_{TE}}} \hat{L}_{TE \nu} (\xi) e^{-i \frac{\delta k}{2} \xi} - \frac{1}{\sqrt{\upsilon_{TM}}} \hat{L}^{\dagger}_{TM (-\nu)} (\xi) e^{i \frac{\delta k}{2} \xi} \right). 
\end{eqnarray}%
Here $D(\nu)$ is the phase mismatch for TE and TM modes at frequencies $\frac{\omega_p}{2} + \nu$ and $\frac{\omega_p}{2} - \nu$, respectively. The square root in Eq.~(\ref{kapa}) should be taken as $\sqrt{Z} = \sqrt{|Z|} e^{i \frac{1}{2} {\rm Arg}[Z]}$. It follows from Eq.~(\ref{sK}) that $K_{+} K_{-} = e^{- 2i {\rm Arg}[g]}$. 
Let's calculate the spectral fluxes of photons at the cross section $z = L$ of the waveguide. We will use the relationship
\begin{equation}
Q_{N}(t,L) = \left\langle \hat{a}_{N}^{\dagger }\left( t,L\right) \hat{a}_{N}\left( t,L\right) \right\rangle = \int_{\Delta \omega} d\nu \left[ \int_{\Delta \omega} d\nu' \left\langle  \hat{a}_{N\nu}^{\dagger }\left( L\right) \hat{a}_{N\nu'}\left( L\right) \right\rangle e^{i (\nu - \nu')t} \right].
\label{sqn}
\end{equation}%

In the absence of coherent incident fields we have $\left\langle  \hat{a}_{N\nu}^{\dagger }\left( z\right) \hat{a}_{N\nu'}\left( z\right) \right\rangle \propto \delta(\nu - \nu')$; therefore 
\begin{equation}
Q_{N}(t,L) = \int_{\Delta \omega} Q_{N\nu}(L) d\nu,
\label{sqn2}
\end{equation}
\begin{equation}
Q_{N\nu}(L) = \int_{\Delta \omega} d\nu' \left\langle  \hat{a}_{N\nu}^{\dagger }\left( L\right) \hat{a}_{N\nu'}\left( L\right) \right\rangle .
\label{sqnu3}
\end{equation}

Using the solution (\ref{ssolution for ceq}) in Eq.~(\ref{sqnu3}) for vacuum boundary conditions given by Eq.~(\ref{sabound}) and Langevin noise given by Eqs.~(\ref{scommutator for L}) and (\ref{scf for L}), we arrive at
\begin{equation}
Q_{N\nu}(L) = Q^{(s)}_{N\nu}(L) + Q_{N\nu}^{\rm noise}(L), 
\label{sqnu4}
\end{equation}
where we separated the ``signal'' component of the flux $Q^{(s)}_{N\nu}$ and the noise component $Q_{N\nu}^{\rm noise}$ which does not depend on the boundary conditions for the fields: 
\begin{equation}
Q^{(s)}_{TE\nu}(L) = Q^{(s)}_{TM(-\nu)}(L) = e^{-\left( 
\frac{\Gamma_{TE}}{\upsilon_{TE}} + \frac{\Gamma_{TM}}{\upsilon_{TM}} \right)L } \frac{|g|^2}{2 \pi} \left| \frac{e^{\kappa L} - e^{-\kappa L}}{2 \kappa} \right|^2 ,  
\label{sqnu5}
\end{equation}
\begin{equation}
    \left( \begin{array}{c}
Q^{\rm noise}_{TE\nu }(L) \\
Q^{\rm noise}_{TM(-\nu) }(L)
\end{array} \right) = 
\left( \begin{array}{c}
\frac{\Gamma_{TM}}{\upsilon_{TM}} \\
\frac{\Gamma_{TE}}{\upsilon_{TE}}
\end{array} \right) \frac{|g|^2}{4\pi |\kappa|^2} F\left(\mu_{\pm},L \right),
\label{sqnoise}
\end{equation}
where 
\begin{eqnarray}
F\left(\mu_{\pm},L \right) &=& \int_0^L \left| e^{\mu_{+} (L - \xi)} - e^{\mu_{-} (L - \xi)}  \right|^2 d\xi 
\nonumber \\
&=& \frac{e^{2{\rm Re}[\mu_{+}]L} - 1}{2 {\rm Re}[\mu_{+}]} + \frac{e^{2{\rm Re}[\mu_{-}]L} - 1}{2 {\rm Re}[\mu_{-}]} - 2 {\rm Re} \left[ \frac{e^{(\mu_{+}^* + \mu_{-})L} - 1}{\mu_{+}^* +\mu_{-}}  \right];
\label{sTheta}
\end{eqnarray}
$2{\rm Re}[\mu_{\pm}] = - \left( \frac{\Gamma_{TE}}{\upsilon_{TE}} + \frac{\Gamma_{TM}}{\upsilon_{TM}} \right) \pm 2 {\rm Re} [\kappa] $,  
$\mu_{+}^* + \mu_{-} = - \left( \frac{\Gamma_{TE}}{\upsilon_{TE}} + \frac{\Gamma_{TM}}{\upsilon_{TM}} \right) - 2 i {\rm Im} [\kappa]$. When calculating the noise components of the fluxes we assumed that at optical frequencies the reservoir can be treated as having zero temperature. In this case we have $n_T(\omega_N) = 0$ in Eq.~(\ref{scf for L}). 

The dynamic components of the fluxes in TE and TM modes are equal to each other even though their absorption losses may be very different; see Eq.~(\ref{sqnu5}) and also Eq.~(\ref{sqtm}) in the non-degenerate case. This property holds only for vacuum boundary conditions with zero average number of photons. For a classical field or any multiquantum field the mode with lower losses will accumulate a higher flux. 

The frequency spectrum of the downconverted photons is determined by the dependence $\kappa(\nu)$ in Eqs.~(\ref{skapa}), (\ref{sdnu}).  
As follows from Eqs.~(\ref{skapa}), (\ref{sdnu}), and (\ref{sqnu5}),  in the absence of dissipation the parametric amplification occurs in the frequency interval
$|D(\nu)| = \left| \delta k + \nu \left( 
\frac{1}{\upsilon_{TE}}- \frac{1}{\upsilon_{TM}} \right) \right| < 2 |g|$. For $D(\nu) \rightarrow 0$ the threshold for parametric amplification is determined by dissipation:  $ \frac{\Gamma_{TE} \Gamma_{TM}}{\upsilon_{TE}\upsilon_{TM}} < |g|^2$. Taking into account Eq.~(\ref{sg}), the last inequality can be written as $\Gamma_{TE} \Gamma_{TM} < \frac{|A|^2}{\hbar^2}$, which is exactly the condition for the parametric decay in the initial-value problem \cite{tokman2019}. 

In the absence of dissipation and detuning, i.e. when $\Gamma_{TE} = \Gamma_{TM} = D(\nu) = 0$, the spatial coefficient of amplification is $|g| = \frac{A}{\hbar \sqrt{\upsilon_{TE}\upsilon_{TM}}}$, whereas the growth rate in time for an associated initial-value problem is $\gamma = \frac{|A|}{\hbar}$; see, e.g., \cite{tokman2019}. The relationship $|g| = \frac{\gamma}{ \sqrt{\upsilon_{TE}\upsilon_{TM}}}$ allows one to express the characteristic time of parametric interaction $t_{int}$ through the parametric amplification length $L_z$ as $ t_{int} = \frac{L_z}{ \sqrt{\upsilon_{TE}\upsilon_{TM}}}$ which was used in Sec.~3 of the main paper. 

As one can see from Eq.~(\ref{sqnoise}), the decay photon fluxes ``swap'' their noise components in the SPDC process: the photon flux in the TE mode is proportional to the absorption coefficient of the TM mode and vice versa. Therefore, when the noise reservoir is at zero temperature, there occurs parametric transfer of quantum noise between the two decay modes while  the photon flux of a given mode does not have any contribution from its own noise component. This feature is characteristic of the down-conversion; one can show that in the up-conversion process the Langevin noise don't make any contribution to the upconverted photon flux as long as the reservoir is at zero temperature. 

It follows from Eqs.~(\ref{sqnu5})-(\ref{sTheta}) that the relative contribution of the Langevin noises is negligible in the parametric amplification regime when $|g| \gg \frac{\Gamma_{TE}}{\upsilon_{TE}}, \frac{\Gamma_{TM}}{\upsilon_{TM}} $. If the parametric gain is low, $|g| \leq \frac{\Gamma_{TE}}{\upsilon_{TE}}, \frac{\Gamma_{TM}}{\upsilon_{TM}} $, the relative contribution of noise is small for short enough waveguide lengths, when $\left( \frac{\Gamma_{TE}}{\upsilon_{TE}}, \frac{\Gamma_{TM}}{\upsilon_{TM}} \right) L \ll 1$. 

The above solution provides analytic expressions for the fluxes and spectra of downconverted photons including the effects of the phase and group mismatch and absorption. We can further simplify these expressions in two limiting cases. \\

{\it (a) Parametric gain much higher than the parametric threshold }\\

In this case $|g| \gg \frac{\Gamma_{TE}}{\upsilon_{TE}}, \frac{\Gamma_{TM}}{\upsilon_{TM}} $ and we obtain
\begin{equation}
Q_{TE\nu}(L) = Q_{TM(-\nu)}(L) \approx \frac{|g|^2 e^{-\left( 
\frac{\Gamma_{TE}}{\upsilon_{TE}} + \frac{\Gamma_{TM}}{\upsilon_{TM}} \right)L } }{2\pi |\kappa|^2} \left\{ 
\begin{array}{c} 
\sinh^2(|\kappa| L) \; {\rm for} \; D(\nu) < 2 |g| \\
\sin^2(|\kappa| L) \; {\rm for} \; D(\nu) > 2 |g|
\end{array}
\right. , 
\label{sqnu33}
\end{equation}
where $|\kappa|^2 \approx \left| |g|^2 - \frac{1}{4} D^2(\nu) \right|$. Clearly, the flux of downconverted photons is nonzero even outside the parametric amplification bandwidth; however, it decays at large detunings as $\frac{1}{|\kappa|^2}$ and gets absorbed at propagation distances larger than the absorption length. \\  

{\it (b) Low parametric gain or high absorption losses }

In this case $|g| \ll \frac{\Gamma_{TE}}{\upsilon_{TE}}, \frac{\Gamma_{TM}}{\upsilon_{TM}} $ and the flux of downconverted photons decays over the distances larger than the absorption length at all frequencies. This is the realistic situation for a laser device, as one can see from the numerical estimates below. The expression for the flux is especially simple for propagation distances shorter than the absorption length, where it becomes
\begin{equation}
Q_{TE\nu}(L) = Q_{TM(-\nu)}(L) = \frac{|g|^2 \sin^2(|\kappa| L)}{2 \pi |\kappa|^2}  , 
\label{sqnu44}
\end{equation}
where $|\kappa| \approx \frac{1}{2} |D(\nu)|$. If, in addition, $|\kappa| L \ll 1$, we obtain an especially simple expression 
\begin{equation}
Q_{TE\nu}(L) = Q_{TM(-\nu)}(L) = \frac{|g|^2 L^2} {2 \pi}  , 
\label{sqnu55}
\end{equation}
valid at $\frac{L}{2} |D(\nu)| \ll 1$. These expressions for spectral flux densities have to be integrated over the bandwidth $\Delta \omega$ determined by the detection system to obtain the total flux.

When the detection bandwidth $\Delta \omega $ is narrow enough, namely $%
\Delta \omega \frac{\left\vert \upsilon_{TE}-\upsilon_{TM}\right\vert }{2\upsilon_{TE}\upsilon_{TM}}\ll
\left\vert g\right\vert \approx $ $\kappa $ (i.e., the detection bandwidth
is narrower than the parametric amplification bandwidth), after neglecting dissipation and noise and assuming $\delta k = 0$ the solution in Eq.~(\ref{ssolution for ceq}) can be easily summed over frequencies. Returning to the operators $\hat{c}_{N}$
 we obtain 
\begin{equation}
\hat{c}_{TE}\left( z,t\right) =\hat{c}_{TE}^{\left( 0\right) }\left( t-\frac{%
\upsilon_{TE}+\upsilon_{TM}}{2\upsilon_{TE}\upsilon_{TM}}z\right) \cosh \left( \kappa z\right)
+ie^{i\phi }\sqrt{\frac{\upsilon_{TM}}{\upsilon_{TE}}}\hat{c}_{TM}^{\left( 0\right)
\dagger }\left( t-\frac{\upsilon_{TE}+\upsilon_{TM}}{2\upsilon_{TE}\upsilon_{TM}}z\right) \sinh \left(
\kappa z\right) ,  \label{scte}
\end{equation}%
\begin{equation}
\hat{c}_{TM}\left( z,t\right) =\hat{c}_{TM}^{\left( 0\right) }\left( t-\frac{%
\upsilon_{TE}+\upsilon_{TM}}{2\upsilon_{TE}\upsilon_{TM}}z\right) \cosh \left( \kappa z\right)
+ie^{i\phi }\sqrt{\frac{\upsilon_{TE}}{\upsilon_{TM}}}\hat{c}_{TE}^{\left( 0\right)
\dagger }\left( t-\frac{\upsilon_{TE}+\upsilon_{TM}}{2\upsilon_{TE}\upsilon_{TM}}z\right) \sinh \left(
\kappa z\right) ,  \label{sctm}
\end{equation}%
where $\phi = {\rm Arg}[g]$.

We can find an exact nonperturbative solution to Eqs.~(\ref{sceq for cte}) and (\ref{sceq for ctm}) for negligible Langevin noise, losses, and phase mismatch $\Gamma_{TE} = \Gamma_{TM} = \delta k = 0$ in the different way if we notice that after the substitution $x = z - \upsilon_{TE} t$ and $y = z - \upsilon_{TM} t$ they are reduced to the hyperbolic equation 
\begin{equation*}
    \left[ \frac{\partial^2}{\partial x \partial y} + \frac{1}{\hbar^2} \frac{|A|^2}{(\upsilon_{TE} - \upsilon_{TM})^2} \right] \Phi = 0,
\end{equation*}
where $\Phi = \hat{c}_{TE},\hat{c}_{TM}^{\dagger}$. Its solution can be written in quadratures following the Riemann-Volterra method; see, e.g., Chapter 10.3-6 in \cite{korn}. However, the solution method based on the Fourier transformation which we used is more convenient in this case. Indeed, it gives us an explicit equation for the SPDC frequency bandwidth and highlights the correlations between the spectral photon fluxes in different frequency bins. Furthermore, the spectral approach provides a straightforward way of including finite phase mismatch, Langevin noise, and absorption losses, see Eqs.~(\ref{ssolution for ceq})-(\ref{skapa}). The spectral method also makes it straightforward to incorporate the quantum dynamics of the reservoir noise (e.g. squeezing) for a finite temperature of the reservoir  \cite{erukhimova2017}. 


\section{ Derivations for Sec.~V of the main paper: the boundary-value problem for the Schr\"{o}dinger equation
}

In the previous section we used space-time-dependent Heisenberg creation and annihilation
operators of the optical fields to solve the boundary-value problem. This
approach is approximate as it assumes that the radiation wavelength is much
shorter than the spatial scale at which the field intensity changes
significantly \cite{fain1969}. For the boundary-value problem, the observables are determined using a constant Heisenberg-picture state vector $\left\vert \Psi
\left( t = 0\right) \right\rangle $ at the boundary $ z = 0$. Within the same approximation one can also
introduce the notion of space-dependent state vector which would be
equivalent to the space evolution of Heisenberg operators. With this goal in
mind, let's look at the equations for the spectral components of the field
operators neglecting for simplicity the Langevin noise and dissipation, and taking $\delta k = 0$. When all three fields participating in the SPDC are quantized, the Heisenberg equations for the operators of the pump and decay fields $\hat{a}_{p\nu}, \hat{a}_{TE\nu }$, and $\hat{a}_{TM\nu}$ take the form
\begin{equation}
-i\frac{\nu }{\upsilon_{TE}}\hat{a}_{TE\nu }+\frac{\partial }{\partial z}\hat{a}%
_{TE\nu }=iG\int \int_{\Delta \omega \Delta \omega }d\nu ^{\prime }d\nu ^{\prime \prime
}\delta \left( \nu +\nu ^{\prime }-\nu ^{\prime \prime }\right) \hat{a}_{p\nu ^{\prime \prime }}\hat{a}_{TM\nu ^{\prime }}^{\dagger },
\label{scoupled eqs 1}
\end{equation}%
\begin{equation}
i\frac{\nu ^{\prime }}{\upsilon_{TM}}\hat{a}_{TM\nu ^{\prime }}^{\dagger }+\frac{%
\partial }{\partial z}\hat{a}_{TM\nu ^{\prime }}^{\dagger }=-iG^{\ast }\int
\int_{\Delta \omega \Delta \omega  }d\nu d\nu ^{\prime \prime }\delta \left( \nu +\nu
^{\prime }-\nu ^{\prime \prime }\right) \hat{a}_{p\nu ^{\prime \prime
}}^{\dagger }\hat{a}_{TE\nu },  \label{scoupled eqs 2}
\end{equation}%
\begin{equation}
-i\frac{\nu ^{\prime \prime }}{\upsilon_{p}}\hat{a}_{p\nu ^{\prime \prime }}+\frac{%
\partial }{\partial z}\hat{a}_{p\nu ^{\prime \prime }}=iG^{\ast }\int
\int_{\Delta \omega \Delta \omega }d\nu d\nu ^{\prime }\delta \left( \nu +\nu ^{\prime
}-\nu ^{\prime \prime }\right) \hat{a}_{TE\nu }\hat{a}_{TM\nu ^{\prime }}.
\label{scoupled eqs 3}
\end{equation}%
These operators satisfy commutation relations (\ref{scr}) in which $N = TE,TM,p$. The overlap integral $G$ is determined in Eq.~(47) of the main paper. 

Equations (\ref{scoupled eqs 1})-(\ref{scoupled eqs 3}) are equivalent to 
\begin{equation}
\frac{\partial }{\partial z}\hat{O}=\frac{i}{\hbar }\left[ \hat{H}_{eff},%
\hat{O}\right] ,  \label{sHeisenberg equation}
\end{equation}%
where $\hat{O}=\hat{a}_{p\nu}, \hat{a}_{TE\nu },\hat{a}_{TM\nu}$ and 
\begin{eqnarray}
\hat{H}_{eff} = 2\pi \hbar \left[ \int_{\Delta \omega} d\nu \frac{\nu }{\upsilon_{TE}} 
\hat{a}_{TE\nu }^{\dagger }\hat{a}_{TE\nu } + \int_{\Delta \omega} d\nu \frac{\nu}{\upsilon_{TM}}
\hat{a}_{TM\nu}^{\dagger }\hat{a}_{TM\nu} +  \int_{\Delta \omega} d\nu \frac{\nu }{\upsilon_{p}} 
\hat{a}_{p\nu }^{\dagger }\hat{a}_{p\nu }   \right. \nonumber \\ \left. 
- \int \int \int_{\Delta \omega \Delta \omega \Delta \omega }d\nu d\nu' d\nu'' \delta\left(\nu +\nu^{\prime} - \nu'' \right) 
 \left( G \hat{a}_{p\nu'' } \hat{a}_{TE\nu }^{\dagger }\hat{a}_{TM\nu ^{\prime }}^{\dagger }+h.c.\right) 
  \right] . 
 \label{sheff}
\end{eqnarray}
Note that the operator $\hat{H}_{eff} $ in Eq.~(\ref{sHeisenberg equation}) generates translations along $z$, not time, and therefore it has the dimension of momentum. As shown in Sec.~5 of the main paper, $\hat{H}_{eff} $ determines the spatial evolution of the state vector,  
\begin{equation}
i\hbar \frac{\partial }{\partial z}\left\vert \Psi \right\rangle =\hat{H}%
_{eff}\left\vert \Psi \right\rangle.  \label{sspace evolution equation}
\end{equation}

To avoid cumbersome derivations, we will switch from the continuous spectrum to a discrete set of frequencies; see, e.g., \cite{mandel}. We consider the parametric decay of a quasi-monochromatic pump mode at frequency $\omega_p$.  The spectrum of signal and idler photons is convenient to represent as a set of discrete spectral lines at frequencies $\frac{\omega_p}{2} + \nu$, where $\nu$ span a discrete set of values symmetric with respect to $\omega_p/2$. In this case Eqs.~(\ref{scoupled eqs 1})-(\ref{scoupled eqs 3}) transform into 
\begin{eqnarray} 
 -i\frac{\nu}{\upsilon_{TE}} \hat{a}_{TE\nu }   +\frac{\partial }{\partial z} \hat{a}_{TE\nu } = iG \hat{a}_p \hat{a}_{TM(-\nu) }^{\dagger }, 
\label{dis1} \\
i\frac{(-\nu) }{\upsilon_{TM}} \hat{a}_{TM(-\nu) }^{\dagger } +\frac{\partial }{\partial z} \hat{a}_{TM(-\nu) }^{\dagger } = - iG^{\ast } \hat{a}_{p}^{\dagger }\hat{a}_{TE\nu }, \label{dis2} \\
\frac{\partial }{\partial z} \hat{a}_{p} =  iG^{\ast } \sum_{\nu} \hat{a}_{TE\nu } \hat{a}_{TM(-\nu) }.   \label{dis3}
\end{eqnarray}
In Eq.~(\ref{dis3}) we assume that $\nu = 0$ is the only option for the pump field and we define $\hat{a}_{p,\nu = 0} = \hat{a}_{p}$.

The transition from Eqs.~(\ref{scoupled eqs 1})-(\ref{scoupled eqs 3}) to Eqs.~(\ref{dis1})-(\ref{dis3}) corresponds to the renormalization of the operators $\hat{a}_{N\nu}$. The quantities  $\left\langle  \hat{a}_{N\nu}^{\dagger } \hat{a}_{N\nu} \right\rangle $ in Eqs.~(\ref{dis1})-(\ref{dis3}) are now the total fluxes of photons of a given polarization within a given spectral line, i.e., they have the dimension of sec$^{-1}$. The operators defined in this way satisfy the commutation relations that follow from Eq.~(\ref{cr}) (see also the Supplemental Material in \cite{tokman2013}): 
\begin{equation}
\left[ \hat{a}_{N\nu },\hat{a}_{N^{\prime }\nu ^{\prime }}^{\dagger }\right]
=  \frac{\Delta \omega}{2 \pi} \delta _{NN^{\prime }}\delta _{\nu \nu ^{\prime }}.  \label{scomrel}
\end{equation}

The Heisenberg equations (\ref{dis1})-(\ref{dis3}) with commutation relations (\ref{scomrel}) correspond to the following discrete version of the effective `` Hamiltonian'' to be used in Eq.~(\ref{sspace evolution equation}): 
\begin{equation}
\hat{H}_{eff} = \frac{2\pi \hbar}{\Delta \omega} \sum_{\nu} \left[  \frac{\nu }{\upsilon_{TE}} 
\hat{a}_{TE\nu }^{\dagger }\hat{a}_{TE\nu } + \frac{\nu}{\upsilon_{TM}}
\hat{a}_{TM\nu}^{\dagger }\hat{a}_{TM\nu} -
 \left( G \hat{a}_{p} \hat{a}_{TE\nu }^{\dagger }\hat{a}_{TM(-\nu)}^{\dagger }+h.c.\right) 
  \right] .  \label{sHeff2}
\end{equation}

As the next step, we need to define how the operators $\hat{a}_{N\nu}$ in Eq.~(\ref{sHeff2}) act on the state vector and clarify the physical meaning of the latter. The commutation relations (\ref{scomrel}) allow us to define standard states of the boson field according to 
\begin{equation}
\sqrt{\frac{2 \pi}{\Delta \omega}}  \hat{a}_{N\nu }\left\vert n_{N\nu }\right\rangle =\sqrt{n_{N\nu }}\left\vert
\left( n-1\right) _{N\nu }\right\rangle ,\ \ \sqrt{\frac{2 \pi}{\Delta \omega}}  \hat{a}_{N\nu }^{\dagger
}\left\vert n_{N\nu }\right\rangle =\sqrt{\left( n+1\right) _{N\nu }}%
\left\vert \left( n+1\right) _{N\nu }\right\rangle . \label{snewcom}
\end{equation}
When the state vector is expressed in terms of these number states as $\Psi_{N\nu} = \sum_n C_{N\nu}^{(n)}(z) |n \rangle$, the quantities $\left|C_{N\nu}^{(n)}(z) \right|^2 $ have the meaning of the probability to detect at the cross section $z$ the flux of photons $\left\langle  \hat{a}_{N\nu}^{\dagger } \hat{a}_{N\nu} \right\rangle = \frac{\Delta \omega}{2 \pi} n$. The quantity $\hbar \omega_N \frac{\Delta \omega}{2 \pi}$ is the energy flux transported by a single photon with waveform of duration $\frac{2 \pi}{\Delta \omega}$. The bandwidth $\Delta \omega$ and the values of the amplitudes $C_{N\nu}^{(n)}(z=0)$ at the boundary are determined by the properties of the field and the waveguide. 

Therefore, within the discrete approach we need to assign a certain spectral bandwidth $\Delta \omega$ to the pump field, which is defined by externally controlled properties of the pump, and to split the spectrum of decay photons into the spectral lines of the same width. We are not considering a rather exotic situation in which the ``allowed'' spectral bands for the signal and idler photons have to be narrower than the pump field bandwidth. 

Consider a parametric decay when the quantum state at the boundary is $\left\vert \Psi \left( 0\right) \right\rangle = \left\vert 1_{p}\right\rangle
\left\vert 0_{TE,TM }\right\rangle $, where $\left\vert 0_{TE,TM }\right\rangle$ is a vacuum state of the signal and idler photons at all frequencies. In this case the solution to Eq.~(\ref{sspace evolution equation}) must have the form
\begin{eqnarray}
\left\vert \Psi \right\rangle &=& C_{p}\left( z\right) \left\vert 1_{p}\right\rangle
\left\vert 0_{TE,TM }\right\rangle \nonumber \\
&+& \sum_{\nu}
C_{W\nu}\left( z\right) \left\vert 0_{p}\right\rangle \left\vert 1_{TE\nu
}\right\rangle \left\vert 1_{TM(-\nu) }\right\rangle \prod_{\nu' \neq \nu, \nu'' \neq -\nu}
\left\vert 0_{TE\nu' }\right\rangle \left\vert 0_{TM\nu''}\right\rangle .  \label{ssol1}
\end{eqnarray}
All other states are forbidden by energy conservation. 

Substituting Eq.~(\ref{ssol1}) into Eq.~(\ref{sspace evolution equation}) with the effective ``Hamiltonian'' (\ref{sHeff2}) leads to the following equations for the coefficients, 
\begin{equation}
\frac{\partial }{\partial z}C_{p}=-i G^* \sqrt{\frac{\Delta \omega}{2\pi}}  \sum_{\nu} C_{W\nu} ,  \label{scp}
\end{equation}%
\begin{equation}
\frac{\partial }{\partial z}C_{W\nu}=i\delta_{\nu} C_{W\nu}- i  G \sqrt{\frac{\Delta \omega}{2\pi}} C_{p},
\label{scwnu}
\end{equation}
where 
\begin{equation}
\delta_{\nu} =\nu \left( \frac{1}{\upsilon_{TM}}-\frac{1}{\upsilon_{TE}} \right).  \label{sdelta}
\end{equation}

Let's assume for simplicity that the nonlinear waveguide allows the parametric decay of the pump into photon pairs within only two symmetric spectral bands $\frac{\omega_p}{2} \pm \nu$, where $\nu$ has only one value. In this case the solution should be sought in the form  
\begin{eqnarray}
\left\vert \Psi \right\rangle &=&C_{p}\left( z\right) \left\vert
1_{p}\right\rangle \left\vert 0_{TE\nu }\right\rangle \left\vert 0_{TM\nu
}\right\rangle \left\vert 0_{TE\left( -\nu \right) }\right\rangle \left\vert
0_{TM\left( -\nu \right) }\right\rangle  \notag \\
&&+C_{W1}\left( z\right) \left\vert 0_{p}\right\rangle \left\vert 1_{TE\nu
}\right\rangle \left\vert 0_{TM\nu }\right\rangle \left\vert 0_{TE\left(
-\nu \right) }\right\rangle \left\vert 1_{TM\left( -\nu \right)
}\right\rangle  \notag \\
&&+C_{W2}\left( z\right) \left\vert 0_{p}\right\rangle \left\vert 0_{TE\nu
}\right\rangle \left\vert 1_{TM\nu }\right\rangle \left\vert 1_{TE\left(
-\nu \right) }\right\rangle \left\vert 0_{TM\left( -\nu \right)
}\right\rangle .  \label{sthe solution}
\end{eqnarray}
Equations (\ref{scp})-(\ref{sdelta})    become 
\begin{equation}
\frac{\partial }{\partial z}C_{p}=-i G^* \sqrt{\frac{\Delta \omega}{2\pi}} \left(
C_{W1}+C_{W2}\right) ,  \label{seq for cd}
\end{equation}%
\begin{equation}
\frac{\partial }{\partial z}C_{W1}=i\delta C_{W1}- i  G \sqrt{\frac{\Delta \omega}{2\pi}} C_{p},
\label{seq for cw1}
\end{equation}%
\begin{equation}
\frac{\partial }{\partial z}C_{W2}=-i\delta C_{W2}-i G \sqrt{\frac{\Delta \omega}{2\pi}} C_{p},
\label{seq for cw2}
\end{equation}%
where
\begin{equation}
\delta =\nu \left( \frac{1}{\upsilon_{TM}}-\frac{1}{\upsilon_{TE}} \right)  \label{sdelta2}
\end{equation}%
and $\nu$ has only one positive value.

The solution for the initial conditions $C_{p}=1$, $C_{W1,2}=0$ is 
\begin{equation}
C_{W1}= -i\frac{G \sqrt{Q_{0}}}{K_{R}}\left[ \sin K_{R}z-i\frac{\delta }{K_{R}}%
\left( \cos K_{R}z-1\right) \right] ,  \label{scw1}
\end{equation}%
\begin{equation}
C_{W2}= -i\frac{G \sqrt{Q_{0}}}{K_{R}}\left[ \sin K_{R}z+i\frac{\delta }{K_{R}}%
\left( \cos K_{R}z-1\right) \right] ,  \label{scw2}
\end{equation}%
\begin{equation}
C_{p}=\frac{\delta ^{2}}{K_{R}^{2}}+\frac{2 Q_{0} \left\vert G\right\vert ^{2}}{%
K_{R}^{2}}\cos K_{R}z  \label{scd}
\end{equation}%
where $Q_0 = \sqrt{\frac{\Delta \omega}{2 \pi}}$ and 
\begin{equation}
K_{R}^{2}=\delta ^{2}+2 Q_{0} \left\vert G\right\vert ^{2},
\label{sRabi wavenumber}
\end{equation}
where $K_R$ is the Rabi wavenumber, introduced in analogy with the Rabi frequency. 

The resulting state vector is
\begin{eqnarray}
\left\vert \Psi \right\rangle  = \left( \frac{\delta ^{2}}{K_{R}^{2}}+\frac{
2 Q_{0} \left\vert G\right\vert ^{2}}{K_{R}^{2}}\cos K_{R}z\right)
\left\vert 1_{p}\right\rangle \left\vert 0_{TE\nu }\right\rangle \left\vert
0_{TM\nu }\right\rangle \left\vert 0_{TE\left( -\nu \right) }\right\rangle
\left\vert 0_{TM\left( -\nu \right) }\right\rangle   \notag \\
-i\frac{G \sqrt{Q_{0}}}{K_{R}}\left[ \sin K_{R}z-i\frac{\delta }{K_{R}}\left(
\cos K_{R}z-1\right) \right] \left\vert 0_{p}\right\rangle \times   \notag \\
\left( \left\vert 1_{TE\nu }\right\rangle \left\vert 0_{TM\nu
}\right\rangle \left\vert 0_{TE\left( -\nu \right) }\right\rangle \left\vert
1_{TM\left( -\nu \right) }\right\rangle +e^{i\varphi _{z}}\left\vert
0_{TE\nu }\right\rangle \left\vert 1_{TM\nu }\right\rangle \left\vert
1_{TE\left( -\nu \right) }\right\rangle \left\vert 0_{TM\left( -\nu \right)
}\right\rangle \right) ,  \label{sstate vector}
\end{eqnarray}%
where%
\begin{equation}
\varphi _{z}=\mathrm{Arg}\left[ \frac{\sin K_{R}z+i\frac{\delta }{K_{R}}%
\left( \cos K_{R}z-1\right) }{\sin K_{R}z-i\frac{\delta }{K_{R}}\left( \cos
K_{R}z-1\right) }\right]   \label{sphase}
\end{equation}

Equation (\ref{sstate vector}) is the generalization of a tripartite entangled state of the  Greenberger-Horne-Zeilinger (GHZ) type  \cite{tokman2021,dur,cunha,shalm,agusti} to the case when the selecton rules and conservation laws allow the decay of an initial excitation of the system into any of the two ``allowed'' boson pairs. 

For small group velocity mismatch, when $\nu \left| \frac{1}{\upsilon_{TM}}-\frac{1}{\upsilon_{TE}} \right| \ll \frac{\sqrt{2 Q_0}\left\vert
G\right\vert }{\hbar }\approx K_{R}$, the quantum state in the waveguide
cross sections defined by $K_{R}z=\frac{\pi }{2}+\pi M$,
\begin{equation}
\left\vert \Psi \right\rangle \approx \left\vert 0_{p}\right\rangle \frac{%
\left\vert 1_{TE\nu }\right\rangle \left\vert 0_{TM\nu }\right\rangle
\left\vert 0_{TE\left( -\nu \right) }\right\rangle \left\vert 1_{TM\left(
-\nu \right) }\right\rangle +\left\vert 0_{TE\nu }\right\rangle \left\vert
1_{TM\nu }\right\rangle \left\vert 1_{TE\left( -\nu \right) }\right\rangle
\left\vert 0_{TM\left( -\nu \right) }\right\rangle }{\sqrt{2}},  \label{sPsi}
\end{equation}
 is one of the entangled Bell states.

Here we considered the parametric decay into two relatively narrow spectral bands.  Now consider a broadband decay in which the total width of the SPDC spectrum is $\Delta \Omega \gg \Delta \omega$. We will use  the method developed in \cite{tokman2021-2} for strong coupling in the systems with inhomogeneous broadening of the spectra.  

The solution to Eq.~(\ref{scwnu}) for the initial conditions $C_{W\nu} = 0$ is
\begin{equation}
C_{W\nu}= - i G \sqrt{\frac{\Delta \omega}{2\pi}} \int_0^z e^{\delta_{\nu} (z - \xi)} C_p(\xi) d\xi. 
\label{sbroad1}
\end{equation}
Substituting it into Eq.~(\ref{scp}) gives 
\begin{equation}
\frac{\partial }{\partial z}C_{p}= - |G|^2 \frac{\Delta \omega}{2\pi}  \sum_{\nu} \int_0^z e^{\delta_{\nu} (z - \xi)} C_p(\xi) d\xi.   \label{scp2}
\end{equation}%
Since the spectrum of decay photons is split into the bands of width $\Delta \omega$, we can transform Eq.~(\ref{scp2}) as 
\begin{equation}
\frac{\partial }{\partial z}C_{p}= - |G|^2 \frac{\Delta \omega}{2\pi}  \sum_{m = -M}^{m = M} \int_0^z e^{\delta_{m} (z - \xi)} C_p(\xi) d\xi,   \label{scp3}
\end{equation}%
where
\begin{equation}
\delta_m = m\Delta \omega \left( \frac{1}{\upsilon_{TM}}-\frac{1}{\upsilon_{TE}} \right) \label{sdelta3}
\end{equation}
and $M = \left[ \frac{\Delta \Omega}{2\Delta \omega} \right]$. Going from summation to integration in Eq.~(\ref{scp3}), 
\begin{equation}
\sum_{m = -M}^{m = M} \int_0^z e^{\delta_{m} (z - \xi)} C_p(\xi) d\xi \Rightarrow \int_{-\Gamma/2}^{\Gamma/2} \frac{d\delta}{\Delta \omega \left| \frac{1}{\upsilon_{TM}}-\frac{1}{\upsilon_{TE}}  \right|} \int_0^z e^{\delta (z - \xi)} C_p(\xi) d\xi,
\label{scp4}
\end{equation}
where
\begin{equation}
   \Gamma =  \Delta \Omega \left| \frac{1}{\upsilon_{TM}}-\frac{1}{\upsilon_{TE}}  \right|, 
\end{equation}
we arrive at 
\begin{equation}
\frac{\partial }{\partial z}C_{p}= - \frac{|G|^2}{\left| \frac{1}{\upsilon_{TM}}-\frac{1}{\upsilon_{TE}}  \right|} \int_0^z d\xi    \left[ \frac{\sin\left[ \frac{\Gamma}{2} (z - \xi) \right]}{\pi (z - \xi)}  \right] C_p(\xi) .   \label{scp5}
\end{equation}

Let's denote by $\lambda$ a typical spatial scale of the function $C_p(\xi)$. If $\frac{\Gamma}{2} \lambda \gg 1$, one can replace $ \frac{\sin\left[ \frac{\Gamma}{2} (z - \xi) \right]}{\pi (z - \xi)} \Rightarrow \delta(z-\xi)$ in Eq.~(\ref{scp5}), which gives
\begin{equation}
\frac{\partial }{\partial z}C_{p}= - \kappa C_p,    \label{scp6}
\end{equation}
where
\begin{equation} 
\kappa = \frac{|G|^2}{\left| \frac{1}{\upsilon_{TM}}-\frac{1}{\upsilon_{TE}}  \right|}.
\end{equation}
For the initial condition $C_p = 1$ we obtain 
\begin{equation}
C_{p}= e^{- \kappa z}.    \label{scp7}
\end{equation}
Equations (\ref{scp6}), (\ref{scp7}) are valid when $\frac{\Gamma}{2 \kappa} \gg 1$,
which corresponds to the condition 
\begin{equation} 
\alpha = \frac{\sqrt{\Delta \Omega}}{|G|}  \left| \frac{1}{\upsilon_{TM}}-\frac{1}{\upsilon_{TE}}  \right| \gg 1.
\label{slimit1}
\end{equation}

In the opposite limit $\frac{\Gamma}{2} \lambda \ll 1$  we will seek the solution of Eq.~(\ref{scp5}) as $C_p \propto e^{qz}$. The right-hand side of Eq.~(\ref{scp5}) can be expanded in powers of $q^{-1}$ by repeated integration by parts. Denoting $ \frac{\sin\left[ \frac{\Gamma}{2} (\xi-z) \right]}{\pi (\xi-z)} = D(\xi - z)$, we obtain 
$$
\int_{-\infty}^{z} d\xi D(\xi - z) e^{q \xi} = \frac{D(0)}{q} + \sum_{n = 1}^{\infty} (-1)^n \left[ \frac{d^n D(\xi - z)}{d\xi^n}\right]_{\xi = z} \frac{1}{q^{n+1}}, 
$$
where 
$$
\left[ \frac{d^{2n+1} D(\xi - z)}{d\xi^{2n+1}}\right]_{\xi = z} = 0; \; \left[ \frac{d^{2n} D(\xi - z)}{d\xi^{2n}}\right]_{\xi = z} = \frac{(-1)^n (2n)! \left( \frac{\Gamma}{2} \right)^{2n+1}}{\pi (2n+1)!}. 
$$
Taking into account that $\frac{|G|^2}{\left| \frac{1}{\upsilon_{TM}}-\frac{1}{\upsilon_{TE}}  \right|} \frac{\Gamma}{2\pi} = |G|^2 \frac{\Delta \Omega}{2\pi}$, we obtain from Eq.~(\ref{scp5}) that
\begin{equation}
    q^2 + |G|^2 \frac{\Delta \Omega}{2\pi} \left( 1 - \sum_{n = 1}^{\infty} \frac{(-1)^n (2n)! }{(2n+1)!} \left( \frac{\Gamma}{2q} \right)^{2n} \right) = 0.
    \label{scp8}
\end{equation}
In zeroth order with respect to a small parameter $\frac{\Gamma}{2q}$ we have 
\begin{equation}
    q^2 + |G|^2 \frac{\Delta \Omega}{2\pi}  = 0; \; C_p \propto e^{\pm  i|G| \sqrt{\frac{\Delta \Omega}{2\pi} } z}.
\end{equation}
For the initial condition $C_p = 1$ this solution corresponds to spatial Rabi oscillations, 
\begin{equation}
    C_p \approx \cos\left(|G| \sqrt{\frac{\Delta \Omega}{2\pi} } z\right).
    \label{srabi1}
\end{equation}
Equation (\ref{srabi1}) is valid when $\frac{\Gamma}{2 |G| \sqrt{\frac{\Delta \Omega}{2\pi} }} \ll 1$ which corresponds to the region of parameters opposite to that of Eq.~(\ref{slimit1}), i.e., $\alpha \ll 1$. 

Let's now discuss the corrections due to higher order terms with respect to a small parameter $\frac{\Gamma}{2q}$. Substituting $q = q_0 + \eta$ into Eq.~(\ref{scp8}), where $q_0 = \pm  i|G| \sqrt{\frac{\Delta \Omega}{2\pi} }$ and $|q_0| \gg |\eta|$, we obtain 
\begin{equation}
   \eta = \displaystyle \frac{|G|^2 \frac{\Delta \Omega}{2\pi} \sum_{n = 1}^{\infty} \frac{(-1)^n (2n)! }{(2n+1)!} \left( \frac{\Gamma}{2q_0} \right)^{2n}   }{2 q_0 + \frac{|G|^2 \frac{\Delta \Omega}{2\pi}}{q_0} \sum_{n = 1}^{\infty} \frac{(-1)^n 2n (2n)! }{(2n+1)!} \left( \frac{\Gamma}{2q_0} \right)^{2n}     }.
    \label{scp9}
\end{equation}
It is easy to see that all terms in the numerator of Eq.~(\ref{scp9}) are real whereas all terms in the denominator are imaginary. Therefore, $\eta$ is imaginary, i.e. it changes the wavenumber but not the decay constant. This is true in any order with respect to $\frac{\Gamma}{2|q_0|}$. 

The amplitudes $C_{W\nu}$ are easily calculated by substituting the expression for $C_p$ from Eq.~(\ref{scp7}) or (\ref{srabi1}) into Eq.~(\ref{sbroad1}).


\end{document}